\documentclass[final,3p,times]{elsarticle}
\usepackage{graphicx}
\usepackage{graphicx,epsf,subfigure,amsmath}

\usepackage{amssymb}
\usepackage{lineno}

\newcommand{\D}{\mathrm{d}} 

\journal{PEPI}

\begin{document}

\begin{frontmatter}

\title{A systematic numerical study of the tidal instability in a rotating triaxial ellipsoid}

\author[label1]{D. C\'ebron\corref{cor1}}
\ead{cebron@irphe.univ-mrs.fr}
\cortext[cor1]{corresponding author}
\author[label1]{M. Le Bars}
\author[label2]{J. Leontini}
\author[label1]{P. Maubert}
\author[label1]{P. Le Gal}

\address[label1]{Institut de Recherche sur les Ph\'enom\`enes Hors Equilibre, UMR 6594, CNRS et Aix-Marseille Universit\'es, 49 rue
F. Joliot-Curie, BP146, 13384 Marseille, France.}
\address[label2]{Department of Mechanical and Aerospace Engineering, Monash University, Clayton Campus
Victoria 3800, Australia}
% \address[label3]{Laboratoire d'Astrophysique de Toulouse-Tarbes, CNRS et Universit\'e de Toulouse, 14 avenue E. Belin,
% 31400 Toulouse, France }

\begin{abstract}
The full non-linear evolution of the tidal instability is studied
numerically in an ellipsoidal fluid domain relevant for planetary
cores applications. Our numerical model, based on a finite element
method, is first validated by reproducing some known analytical results. 
This model is then used to address open questions that were up to now 
inaccessible using theoretical and experimental approaches. Growth rates and 
mode selection of the instability are systematically studied as a 
function of the aspect ratio of the ellipsoid and as a function of 
the inclination of the rotation axis compared to the deformation plane. We 
also quantify the saturation amplitude of the flow driven by the instability and calculate 
the viscous dissipation that it causes. This tidal dissipation can be of major 
importance for some geophysical situations and we thus derive general scaling 
laws which are applied to typical planetary cores.
\end{abstract}


\end{frontmatter}

% \linenumbers

%% main text
\section{Introduction} \label{intro}
Tides are large scale deformations induced by gravitational
interactions that affect all the layers of a given planet or star. They play an
important role in geo- and astrophysics and so, many studies are
devoted to this subject. Their most obvious phenomena are of course
the oceanic flows on Earth, but tides are also responsible for the intense
volcanism on Io for instance. In
stars but also in liquid planetary cores, tidal forcing induces an
elliptical deformation of the rotating streamlines that may excite a
parametric resonance of inertial waves called the elliptical or tidal
instability. This instability could, for instance, be responsible for the 
surprising magnetic field in Io \cite[][]{KerswellMalkus,Lacaze,
herreman} and for fluctuations in the Earth's magnetic field on a
typical timescale of 10,000 years \cite[][]{Aldridge}. It may also
have a significant influence on the orbital evolution of binary
stars \cite[][]{Rieutord} and moon-planet systems
\cite[][]{LeBars09}.

As described in the review of \cite{kerswell_2002}, the
elliptical instability is a three-dimensional instability which may
grow as soon as a rotating fluid possesses elliptical streamlines. As previously stated, it is due to a triadic parametric resonance between two inertial waves of
the rotating fluid \cite[][]{Kelvin_1880} and the underlying strain
field responsible for the elliptic deformation
\cite[][]{Bayly_1986,waleffe_1990}. Such an instability has been
found in many different contexts, including those where the strain field is due to vortex interactions or to elliptically deformed boundaries. It has thus been
the focus of numerous theoretical and experimental studies, devoted
to the dynamics of two-dimensional turbulent flows (e.g.
\cite{Widnall_1974,Moore_1975} or \cite{Haj} for a viscoelastic
fluid), to the stability of wakes \cite[][]{Leweke98_a}, to the
dynamics of vortex pairs
\cite[][]{Leweke98_b,LeDizes_2002,Meunier_2002}, and to the flow
inside deformed rotating cylinders \cite[][]{Gledzer_1975,malkus89,
Eloy_2000,Eloy_2003}, ellipsoids \cite[][]{Gledzer77,Kerswell_1994,lacaze_2004} and shells
\cite[][]{Aldridge,Seyed_2004,lacaze_2005}. More recently,
magnetohydrodynamical effects of the elliptical instability have
been investigated, with applications in MHD turbulence
\cite[][]{thess} or in induction processes in planets
\cite[][]{Lacaze,herreman}.

From a numerical point of view, most studies have been devoted to
the dynamics of deformed two-dimensional vortices in
three-dimensional domains
\cite[e.g.][]{lundgren,Sipp_1998,Lacaze_2007,Roy}. For instance, the
seminal paper of \cite{pierrehumbert} considered the growth of a
perturbation on the planar velocity field associated with an
elliptical vortex inside a box with zero normal-flow boundary
conditions, assuming periodicity in the axial direction; the
linear eigenvalue problem was then solved by spectral methods, and
the growth rate of the elliptical instability was found. Numerical
studies in rotating containers are less numerous. \cite{mason} used a
non-orthogonal elliptico-polar coordinate system in order to solve
the flow in an elliptically deformed cylinder by spectral methods;
the non-linear temporal evolution of two different modes of the
elliptical instability was calculated with no-slip boundary
conditions on the sidewalls and stress-free conditions on the top and the
bottom of the cylinder. \cite{Seyed_2000} determined numerically the
frequencies and the growth rates of the instability in both
ellipsoid and ellipsoidal shell geometries using a linear Galerkin
method, projecting the flow on a selected number of inertial waves.
Finally, \cite{Shangli} studied the stability of self-gravitating
compressible ellipsoidal fluid configurations and pointed out the
occurence of the elliptical instability. However, to the best of our
knowledge, the full non-linear evolution of the tidal instability in
an ellipsoidal geometry has not yet been simulated numerically. This
is the purpose of the present work, which aims at completing our
previous theoretical and experimental investigations
\cite[][]{lacaze_2004,Lacaze,herreman,LeBars09} in accessing global
quantities of fundamental importance for planetary applications.

In this paper, we focus on hydrodynamics only, leaving
thermal and magnetic field interactions with the tidal instability for further
investigations \cite[e.g.][]{Cebron_2010}. In section \ref{section1}, the system and
the numerical method are presented. The numerical model is
validated by comparison with the theoretical results found in the
literature, regarding the flow in a rotating 
sphere submitted to a small and fixed elliptical deformation. In 
section \ref{complexities}, we then study the influence of three 
additional complications of fundamental importance for geophysical
applications: (i) the influence of the length of the
ellipsoid along its rotation axis compared to the mean equatorial
radius (i.e. the body oblateness in geophysical terms), (ii) the
influence of a background rotation, corresponding to the orbital
motion of the companion body responsible for the tidal deformation,
and (iii) the influence of the inclination of the rotation axis
compared to the deformation plane (i.e. the obliquity). In section
\ref{quantif}, we quantify systematically two large-scale quantities
relevant for planetary applications, namely (i) the amplitude of the
instability at saturation, and (ii) the power dissipated by the
instability. General laws in terms of dimensionless numbers are
derived and applied to typical planetary cores in section
\ref{discussion}.

\section{Numerical model and its validation} \label{section1}

\subsection{Definition of the system and description of the numerical method}

\begin{figure}
  \begin{center}
    \epsfysize=7.0cm
    \leavevmode
    \epsfbox{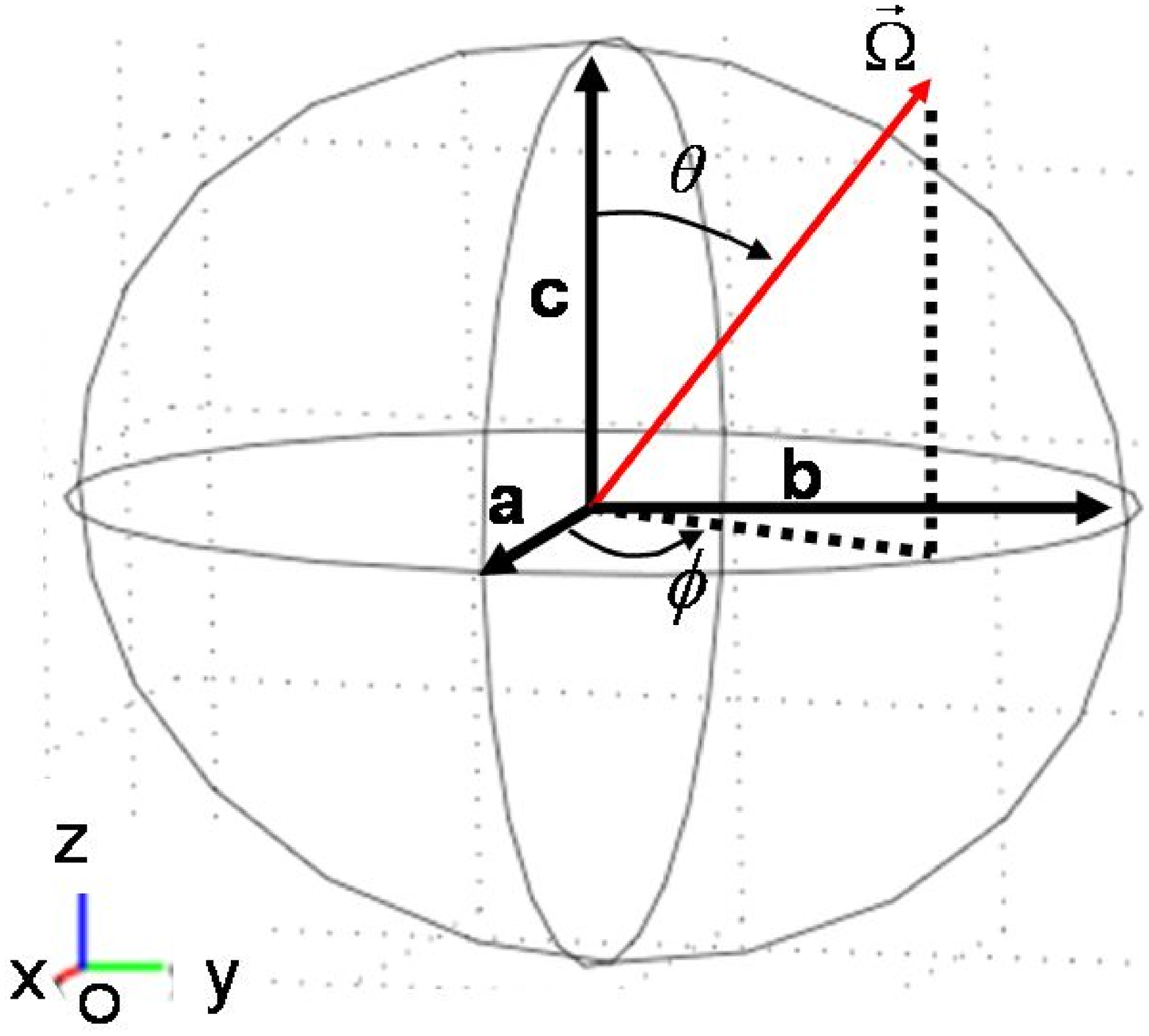}
    \caption{\it{Sketch of the problem under consideration :  a rotating flow inside a triaxial ellipsoid of axes $(a,b,c)$ related to a frame $(Ox,Oy,Oz)$  with a constant tangential velocity all along the boundary in each plane perpendicular to the rotation vector $\boldsymbol{\Omega}$.}}
    \label{cebronfig0}
  \end{center}
\end{figure}

The present study takes place in direct continuity of our
experimental studies of the elliptical instability in a deformed
spheroid \cite[][]{lacaze_2004,Lacaze,herreman,LeBars09}. In these experiments, a
hollow sphere of radius R, molded in a silicone cylinder, is filled
with liquid and set in rotation at a constant angular velocity
$\Omega$ about its axis $(Oz)$, while it is slightly compressed
by a quantity $s$ along the axis $(Ox)$, perpendicular to the
rotation axis. The geometry is then a triaxial ellipsoid of axes
$(a,b,c)=(R-s,R+s,R)$, with an equatorial ellipticity
$\varepsilon=(b^2-a^2)/(a^2+b^2)$ and a constant tangential
velocity along the deformed boundaries, equal to
$\Omega R$ at the equator. Such a configuration is a model for a
liquid planetary core with no solid inner core (e.g. like the jovian
moon Io or the early Earth), surrounded by a deformable mantle with
a constant tangential velocity. Similarly, 
our numerical model studies the
rotating flow inside an ellipsoid of axes $(a,b,c)$ related to the 
frame $(Ox,Oy,Oz)$, with a
constant tangential velocity all along the boundary in each plane
perpendicular to the rotation vector $\boldsymbol{\Omega}$ (see the sketch in figure \ref{cebronfig0}). In order to extend our previous
experimental study, the length $c$ of the ellipsoid can be chosen
independently of the other lengths $a$ and $b$ (with $b>a$). Moreover, the rotation axis of
the ellipsoid can be inclined compared to the c-axis.
Note however that in this paper, unless otherwise specified, we
choose $c$ equal to the mean equatorial radius $R_{eq}=(a+b)/2$ and 
a rotation axis along $(Oz)$, as in the experimental setup. In all
simulations, the fluid is initially at rest and we suddenly impose
at time $t=0$ a constant angular rate $\Omega$ such that the tangential 
velocity along the deformed boundaries in each plane perpendicular 
to the rotation axis, is equal to $\Omega\ (a'+b')/2$, where 
$a'$ and $b'$ are the axes of the elliptic boundary in this plane. 
In the following, results are non-dimensionalised using the mean 
equatorial radius $R_{eq}$ as the length scale and $\Omega^{-1}$ 
as the time scale. Then five dimensionless numbers are used to 
fully describe the system: the Ekman number 
$E=\nu/(\Omega\ R_{eq}^2)$, where $\nu$ is the kinematic viscosity of the fluid, 
the ellipticity $\varepsilon=(b^2-a^2)/(a^2+b^2)$ of the elliptical deformation, the aspect ratio $c/b$ which
quantifies the oblateness of the ellipsoid, and finally the
inclination $\theta$ and declination $\phi$ of the rotation axis. The problem numerically solved is then described by the following system of
dimensionless equations:
\begin{eqnarray}
\frac{\partial \mathbf{u}}{\partial t}+ \mathbf{u} \cdot
\mathbf{\nabla} \mathbf{u} = -\mathbf{\nabla} p + E\
\boldsymbol{\bigtriangleup} \mathbf{u} - 2\ \boldsymbol{\Omega_c^*} \times \boldsymbol{u},
\end{eqnarray}
\begin{eqnarray}
\mathbf{\nabla}  \cdot \mathbf{u} =0.
\end{eqnarray}
where the no-slip boundary conditions are used for the fluid. Note that we 
work is the reference frame where the ellipsoidal shape is at rest, which is 
the inertial frame of reference in most of the paper. The Coriolis force $- 
2\ \boldsymbol{\Omega_c^*} \times \boldsymbol{u}$ is only used in section 
\ref{sec:background} where the whole triaxial ellipsoid is submitted to a 
global rotation at $\Omega_c^*\ \boldsymbol{e_z}$.

Usually, numerical studies of planetary cores benefit from their
spherical geometry to use fast and precise spectral methods. In our
case however, there is no simple symmetry. Our computations are thus performed 
with a standard finite element method widely used in engineering studies, 
which allows to deal with complex geometries, such as our triaxial ellipsoid, 
and to simply impose the boundary conditions. Note that a very efficient  
finite element method was recently introduced by \cite{Chan_2010}, but it is 
up to now restricted to spheroidal geometries ($a=b$). Using a commercial 
software\footnote{COMSOL Multiphysics\textsuperscript{\circledR}}, 
an unstructured mesh with tetrahedral elements was
created. The mesh element type employed is the standard Lagrange element
$P1-P2$, which is linear for the pressure field but quadratic for
the velocity field. Note that no stabilization techniques have been
used in this work. We use the so-called Implicit Differential-Algebraic solver 
(IDA solver), based on backward differencing
formulas \cite[][]{Hindmarsh_2005}. At each time step the system is solved with the sparse
direct linear solver PARDISO\footnote{www.pardiso-project.org}.

\begin{figure}                   % Chaque figure doit avoir pour nom nomfig1.eps,
  \begin{center}
    \begin{tabular}{ccc}
      \setlength{\epsfysize}{6.5cm}
      \subfigure[]{\epsfbox{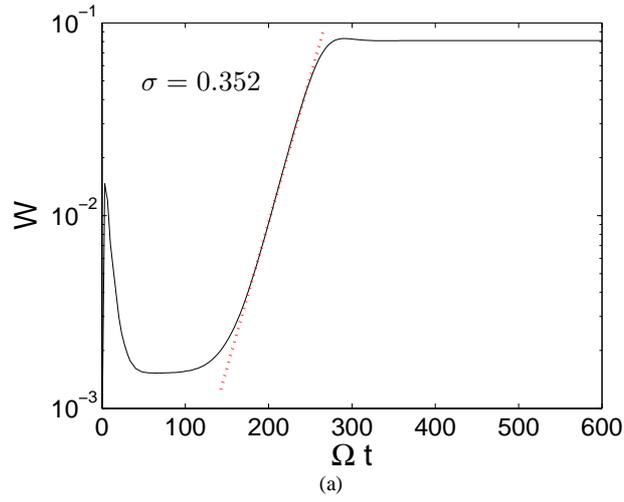}} \\
      \setlength{\epsfysize}{6.5cm}
      \subfigure[]{\epsfbox{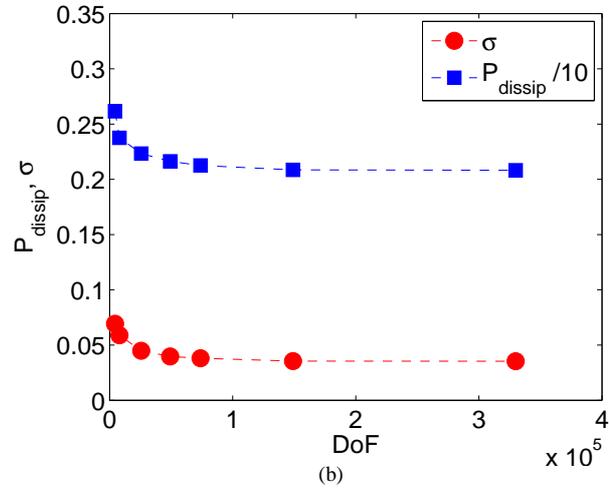}}
    \end{tabular}
    \caption{\it{Simulations performed at $E=1/500$ and $\varepsilon=0.317 $ for $c=\frac{a+b}{2}$.
(a) Time evolution of the mean value of the vertical velocity,
showing the spin-up phase ($\Omega\ t_{E} \approx 22$), the
exponential growth of the tidal instability (see also in dotted line
the exponential fit with a growth rate $\sigma = 0.352$) and its
saturation. (b) Convergence with the number of degrees of freedom
(DoF) of the growth rate $\sigma$ and of the dissipated
power $P_{dissip}$ at saturation. Results presented in (a) are computed for 42459 DoF.}}
    \label{cebronfig1}             % Pensez � mettre le nom du premier auteur � la place de Nom
  \end{center}
\end{figure}

The elliptical instability induces a three-dimensional
destabilization of the initial two-dimensional elliptical
streamlines. To study its global properties, it thus seems natural
to introduce the mean value of the vertical velocity $W=\frac{1}{V}
\iiint_{V} |w|\ \D \tau $, with $w$ the dimensionless vertical velocity
and $V$ the volume of the ellipsoid. The typical evolution of W as a
function of time is shown in figure \ref{cebronfig1} (a) for
$E=1/500$ and $\varepsilon=0.317$. At $t=0$, the
fluid is at rest in the ellipsoid, and the no-slip condition
at the boundary proceeds to set the fluid in rotation. The first peak in W, just after $t=0$, is
due to the Ekman pumping which appears during the spin-up stage,
which typically takes place over the Ekman time $t_{E} =
E^{-1/2}\ {\Omega}^{-1}$, much faster than the viscous time scale
$t_v=R_{eq}^2/ \nu =E^{-1}\ {\Omega}^{-1}$ \cite[][]{benton}. For
instance, in the case shown on figure \ref{cebronfig1} (a), the
dimensionless Ekman time gives $\Omega\ t_{E} \approx 22$, which agrees with the numerical results. After this initial stage, the fluid is essentially in solid body rotation. From this state, the exponential growth of the instability can be seen, before an overshoot and a stationary saturation. Having defined the growth rate $\sigma$ of the elliptical instability as the time constant of the exponential growth, a convergence study on this growth 
rate is given in the figure \ref{cebronfig1} (b). The number of degrees 
of freedom (DoF) used in most of the simulations of this
work ranges between $4\cdot10^4$ DoF and $7\cdot10^4$ DoF, depending
on the ellipticity and the Ekman number, in order to reach a
compromise between a good convergence (see figure \ref{cebronfig1} (b)) and a reasonable CPU time.

\subsection{Validation of our numerical simulations}\label{valid}

\begin{figure}                  % Chaque figure doit avoir pour nom nomfig1.eps,
                                        % ou Nom est celui du premier auteur nomfig2.eps
  \begin{center}
    \begin{tabular}{ccc}
      \setlength{\epsfysize}{5.8cm}
      \subfigure[]{\epsfbox{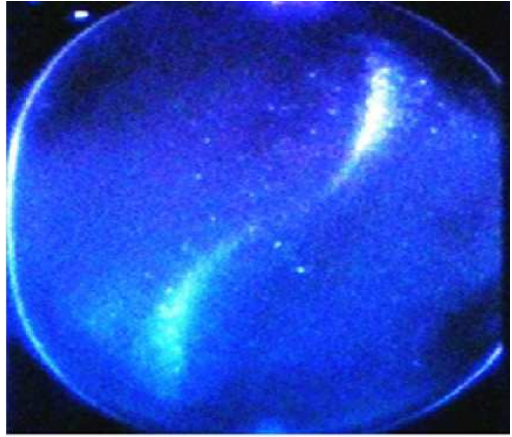}} \\
      \setlength{\epsfysize}{7cm}
      \subfigure[]{\epsfbox{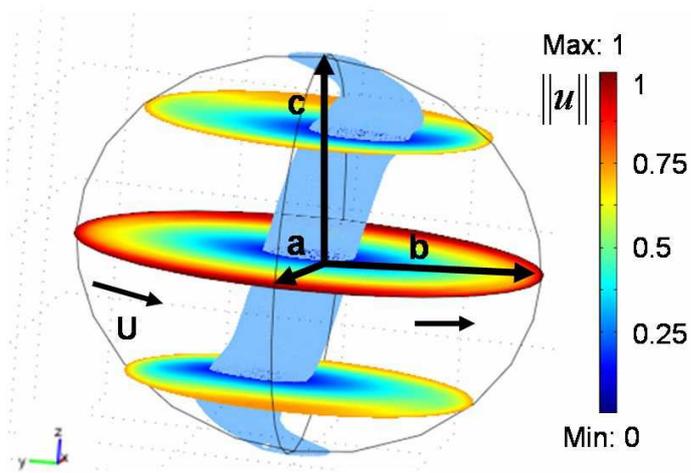}}
     \end{tabular}
    \caption{\it{Validation of the numerical simulations.
(a) Kalliroscopic visualization of the spin-over mode in the
meridional plane of maximum shear for $E=1/4000$ and $\varepsilon=0.16$.
The typical S shape of the rotation axis is due to the combination
of the main rotation imposed by the boundary and the spin-over mode.
(b) Slices of the velocity field $||\bf{u}||$ and surface of
iso-value $||\mathbf{u}||=0.15$ at saturation of the tidal
instability for $E=1/344$, $\varepsilon=0.317$, $c=\frac{a+b}{2},$ and
49900 DoF. The classical S-shape of the spin-over mode is recovered.
}}
    \label{cebronfig2}             % Pensez � mettre le nom du premier auteur � la place de Nom
  \end{center}
\end{figure}

\begin{figure}
  \begin{center}
    \begin{tabular}{ccc}
      \setlength{\epsfysize}{7.5cm}
      \subfigure[]{\epsfbox{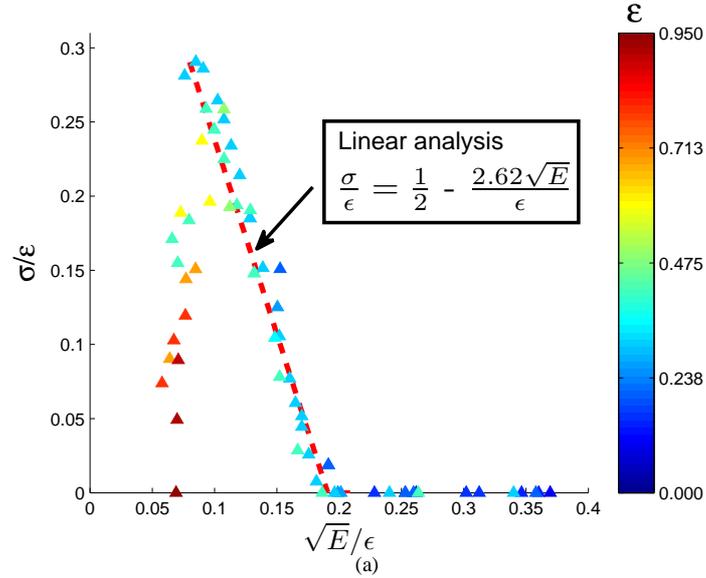}} \\
      \setlength{\epsfysize}{6cm}
      \subfigure[]{\epsfbox{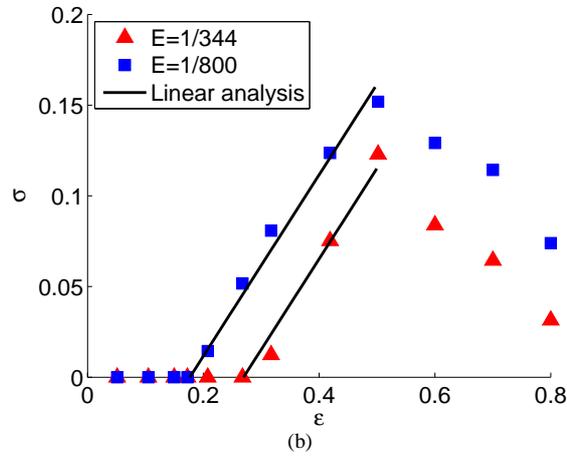}}
    \end{tabular}
    \caption{\it{Validation of the numerical simulations.
(a) Evolution of the growth rate for ($ E \geq 1/2000$, $\varepsilon \leq 0.95$, $c=\frac{a+b}{2} $) and comparison with the
linear theory indicated by a dashed line. The coefficient $2.62$
comes from \cite{lacaze_2004} and is valid in the limit of small
ellipticity. Good agreement is found for values of $\varepsilon$ up
to $0.5$. (b) Evolution of the growth rate depending on the
ellipticity for two values of the Ekman number $E=1/344$ and
$E=1/800$ ($\varepsilon \leq 0.8$, $c=\frac{a+b}{2} $). As
also seen in (a), the growth rate agrees with the linear analytical analysis close to the threshold and then decreases for large values of the
ellipticity.}}
    \label{cebronfig2bis}             % Pensez � mettre le nom du premier auteur � la place de Nom
  \end{center}
\end{figure}

A first visual validation is done on the shape of the flow in
comparing the experimental visualisation on figure \ref{cebronfig2}
(a) with the numerical simulation on figure \ref{cebronfig2} (b). As
can be seen, we recover the classical S-shape of the spin-over mode, which induces
an additional solid body rotation of the fluid around the axis of
maximum strain, i.e. perpendicular to the imposed rotation axis $(Oz)$ and
at an angle of about $45^o$ compared to the deformation axis $(Ox)$ \cite[][]{lacaze_2004}. In
order to quantitatively validate the numerical model, the evolution
of the growth rate of the instability is compared in figure
\ref{cebronfig2bis} to the linear theory given in \cite{lacaze_2004}
for small ellipticities :
\begin{eqnarray}
  \frac{\sigma}{\varepsilon}=\frac{1}{2} - K\ \frac{\sqrt{E}}{\varepsilon},\label{eq:sigmasimple}
\end{eqnarray}
where $K$ is a constant equal to $K=2.62$ in the limit of small
$\varepsilon$ (\cite{Hollerbach_1995,lacaze_2004}). Note that the second term on the right hand side of the 
equation \ref{eq:sigmasimple} corresponds to the viscous damping of the growth rate due to the presence of Ekman layers near boundaries 
\cite[see ][]{Kudlick, Hollerbach_1995}. The expression (\ref{eq:sigmasimple}) allows to define a critical Ekman number for the onset of the instability equal to 
\begin{eqnarray}
  E_c=\left( \frac{\varepsilon}{2\ K} \right)^2 . \label{eq:onset}
\end{eqnarray}
Close to the threshold, the
numerical results closely follow the linear analysis, for values of
ellipticity as large as $\varepsilon \approx 0.5$. Note in particular
that all curves for various $(\varepsilon,E)$ superimpose
providing that $\sigma / \varepsilon$ is expressed as a function of
$\sqrt{E}/\varepsilon$. This is especially interesting in order to
apply our results to planets, whose very small $E$ are not
directly accessible by our numerical tool, but can be compensated for by large
$\varepsilon$ in simulations.

Finally, for very large ellipticities ($\varepsilon>0.5$), figure 
\ref{cebronfig2bis} shows that the growth rate decreases toward 
zero. The elliptical instability finally disappears for ellipticities greater 
than a critical value which depends on the Ekman number. The position of the maximum for the variation of $\sigma$ with the
ellipticity is around $\varepsilon=0.5$, whatever the Ekman
number is. This indicates that the decrease is
probably due to geometrical effects related to the large value of
$\varepsilon$, rather than to any viscous attenuation.

\section{Systematic numerical studies of geophysical complexities}\label{complexities}

\subsection{Influence of the length of the ellipsoid along the rotation axis}\label{influence_length}

Having validated our numerical code, we can investigate the influence of the
aspect ratio $c/b$ on the instability for a given ellipticity in
the equatorial plane and a given rotation rate around the axis
$(Oz)$. This is directly related to geo- and astro-physical
flows as oblateness of planets, like the Earth for instance, is most of the time much stronger than its tidal deformation, meaning that the rotation axis is also the
smallest one. The situation is even more pronounced in certain stars, as for
instance Regulus A whose diameter is about $32 \%$ greater at the
equator than the distance between its poles \cite[][]{mac_2005}.

\begin{figure}                   % Chaque figure doit avoir pour nom nomfig1.eps,
  \begin{center}
    \begin{tabular}{ccc}
      \setlength{\epsfysize}{5.5cm}
      \subfigure[]{\epsfbox{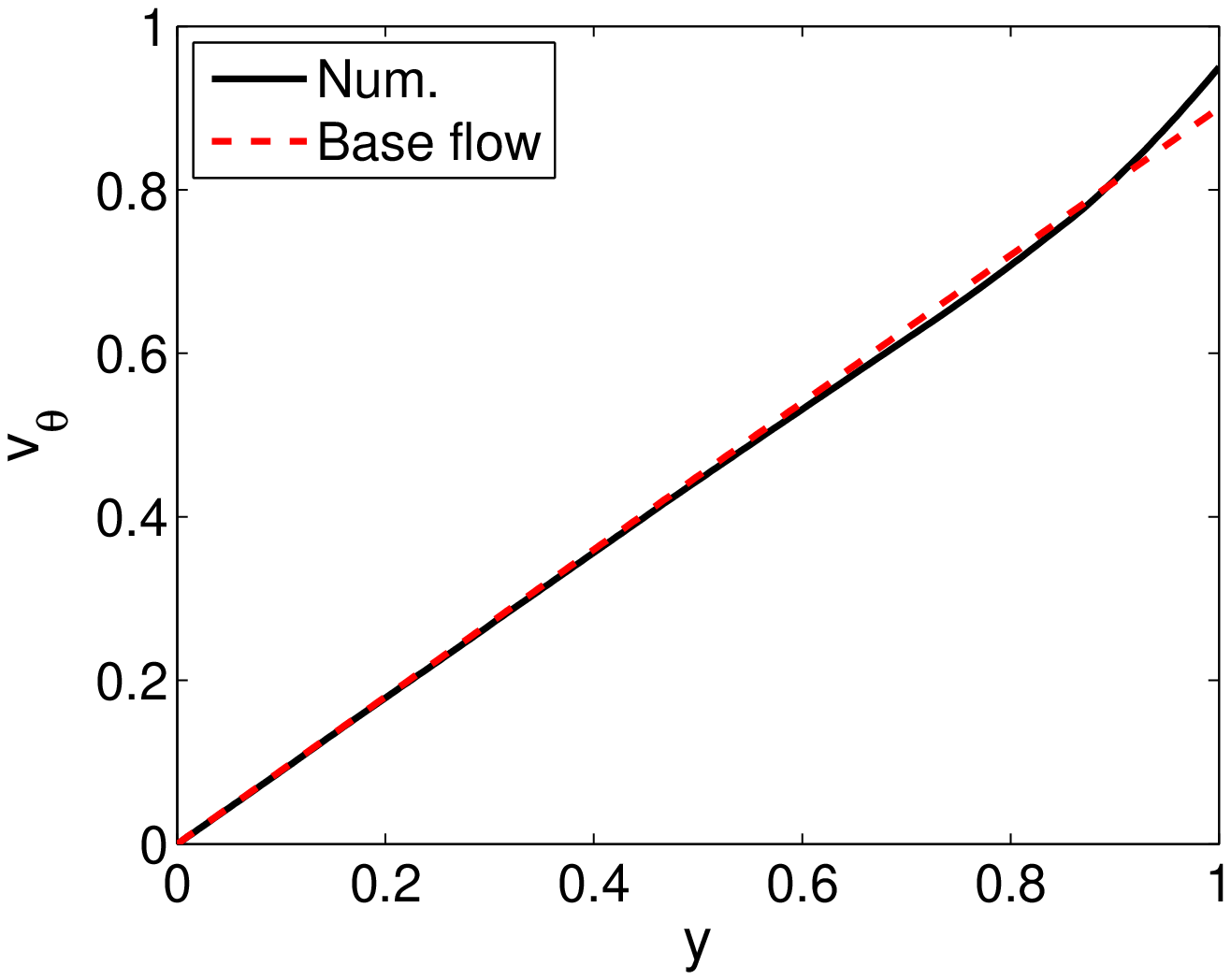}}
      \setlength{\epsfysize}{5.5cm}
      \subfigure[]{\epsfbox{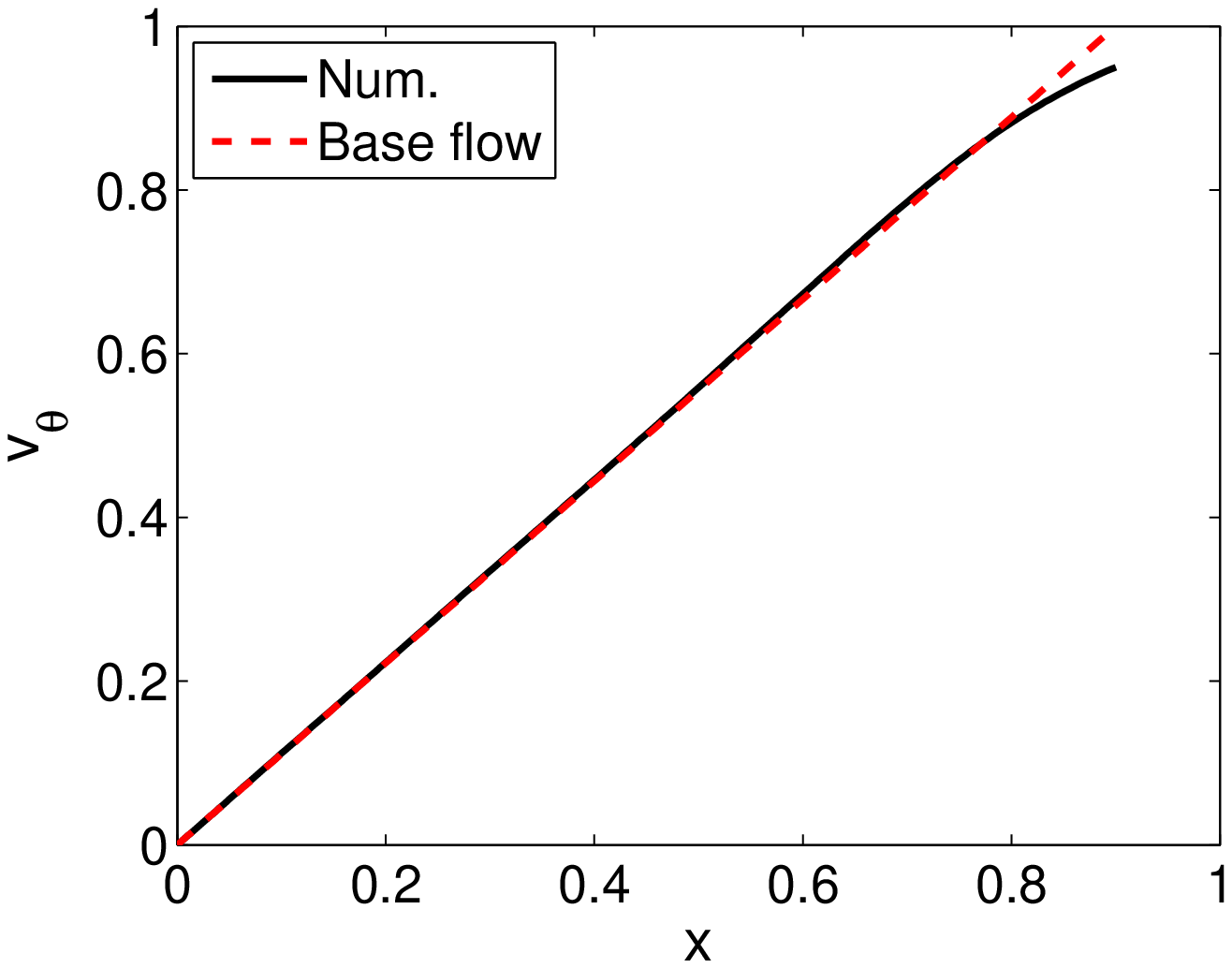}} \\
      \setlength{\epsfysize}{6.5cm}
      \subfigure[]{\epsfbox{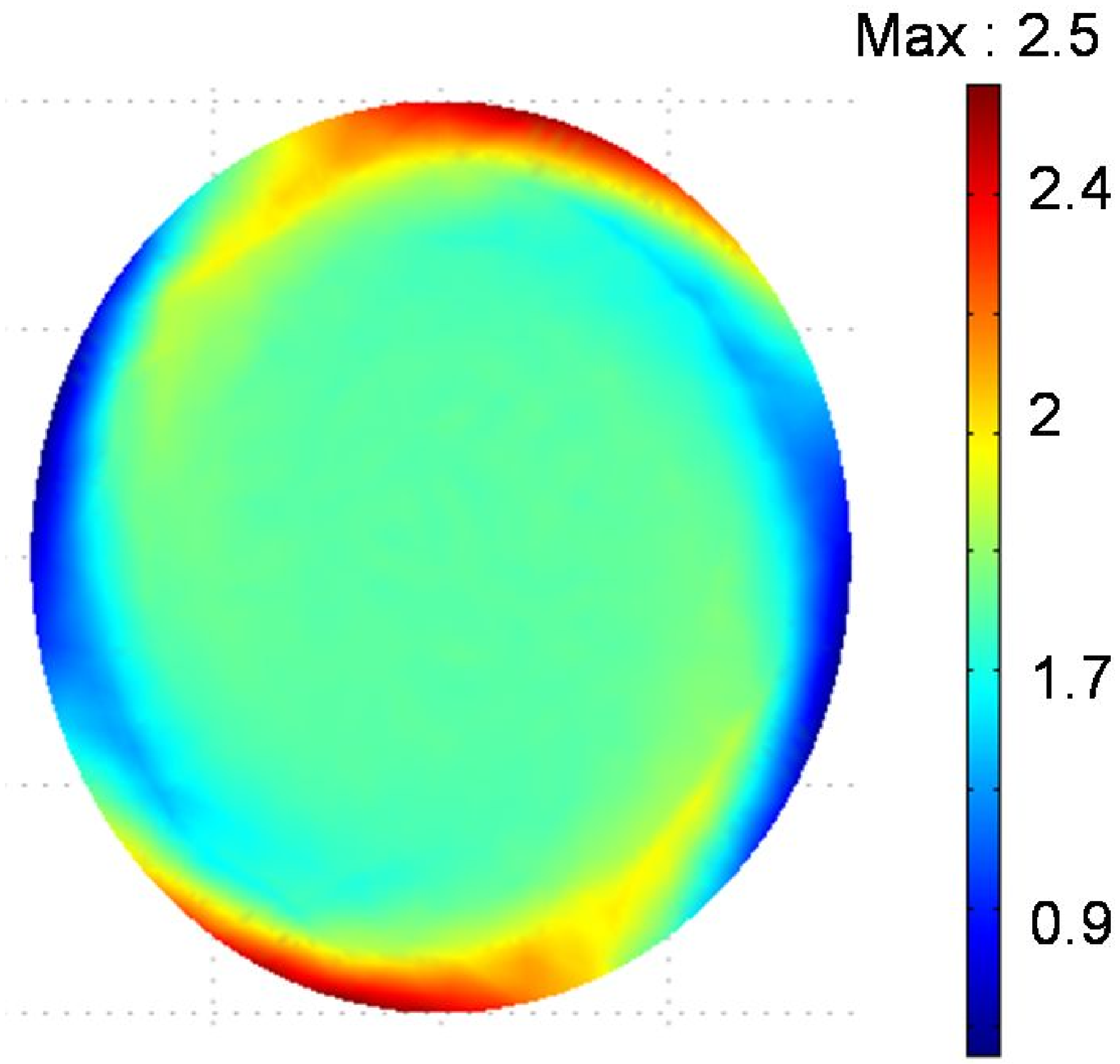}}
   \end{tabular}
    \caption{\it{Azimuthal velocity $v_{\theta}$ for $E=1/95$, $\varepsilon=0.1$,
    $c=\frac{a+b}{2}$. (a) Slice along $(Oy)$ at $x=0$. (b) Slice along $(Ox)$ at $y=0$. (c) Slice in the
equatorial plane of the vertical component of the vorticity. Dashed
lines in (a) and (b) correspond to the theoretical base flow (\ref{baseflow}),
which presents a variable tangential velocity along elliptical
streamlines. Good agreement is found with the numerical results,
except in the small outer viscous boundary layer, where
recirculation cells take place to match the imposed constant
velocity along the boundary. Note in particular that the maximum
velocity is not reached at the boundary but within this viscous
boundary layer.}}
    \label{cebronfig6}             % Pensez � mettre le nom du premier auteur � la place de Nom
  \end{center}
\end{figure}

\begin{figure}                    % Chaque figure doit avoir pour nom nomfig1.eps,
  \begin{center}
    \begin{tabular}{ccc}
      \setlength{\epsfysize}{6.5cm}
      \subfigure[]{\epsfbox{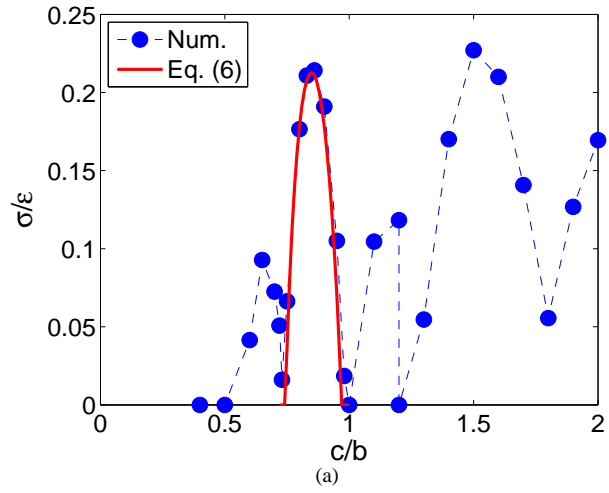}} \\
      \setlength{\epsfysize}{6.5cm}
 %     \subfigure[]{\epsfbox{cebronfig77.eps}} &
%      \setlength{\epsfxsize}{3.0cm}
%      \setlength{\epsfysize}{2.5cm}

      \subfigure[]{\epsfbox{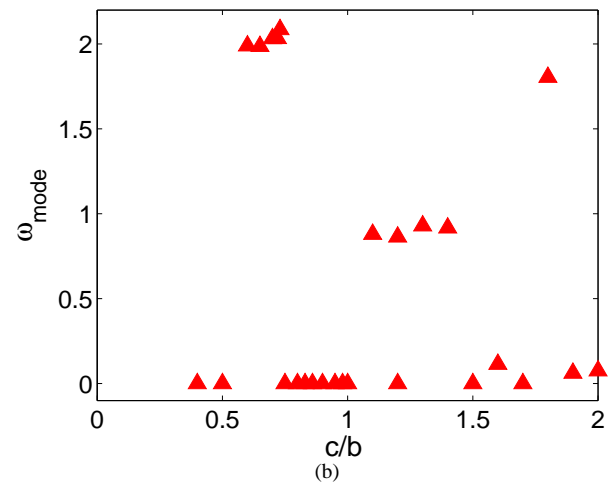}}
    \end{tabular}
    \caption{\it{Influence of the length of the ellipsoid along the rotation axis on the mode selection of the tidal instability ($E=1/688, \varepsilon=0.317 $).
    (a) Variation of the growth rate. (b) Variation of the main frequency of the selected mode determined by Fourier analysis of its saturation state.}}
    \label{cebronfig3}             % Pensez � mettre le nom du premier auteur � la place de Nom
  \end{center}
\end{figure}

Early theoretical work on this aspect was done by
\cite{Kerswell_1994}, who considered the inertial wave basis of an
oblate spheroid $(a=b)$ and calculated the first 60 subharmonic exact
resonances and their growth rate for small ellipticities depending
on the value of $c$ \cite[see also][for a special application to the
case of Io]{KerswellMalkus}. An explicit theoretical answer for the
growth rate has also been found by \cite{Gledzer92}, starting from
the base flow
\begin{eqnarray}
\mathbf{u_b} = -\frac{a}{b}\ y\ \mathbf{e_x} + \frac{b}{a}\ x\ \mathbf{e_y},\label{baseflow}
\end{eqnarray}
and looking for inviscid perturbations
that are linear in space variables, corresponding to the classical
spin-over mode. Here, $\mathbf{e_x}$ and $\mathbf{e_y}$ are respectively the unit
vectors of $(Ox)$ and $(Oy)$. In an open domain, the base flow (\ref{baseflow})
corresponds to elliptical streamlines with an ellipticity
$\varepsilon=(b^2-a^2)/(b^2+a^2)$ as in our numerical model, but
with a variable tangential velocity. As shown in
figure \ref{cebronfig6}, this base flow is a very good approximation
of our configuration where we impose a constant tangential velocity
on the elliptical boundary, outside a small boundary layer close
to the external wall where recirculation cells take place. For such
a flow, the inviscid growth rate determined by \cite{Gledzer92} is:
\begin{eqnarray}
  \sigma=\sqrt{\frac{(b^2-c^2)(c^2-a^2)}{(b^2+c^2)(a^2+c^2)}}\label{eq:sigma_c}
\end{eqnarray}
Note that this theoretical growth rate is valid for $a \leq c \leq
b$ only and is zero for $c=b$ or $c=a$. Then the maximal
theoretical growth rate $\sigma_{\max}=\frac{b-a}{a+b} $ is obtained
for $c= \sqrt{ab}$. One can notice that expression
(\ref{eq:sigmasimple}) leading to $\sigma=\frac{\varepsilon}{2}$ in the
inviscid case is recovered with $a=R_{eq}+s$ and $b=R_{eq}-s$ in the 
limit $\varepsilon \rightarrow 0$. Note also that the experimental choice $c=(a+b)/2$,
equivalent to $c= \sqrt{ab}$ for small ellipticities, results in the growth rate remaining very close to the maximum; 
for instance, even for an ellipticity of $0.7$,
the difference between the growth rate calculated for $c=\frac{a+b}{2}$ and the maximum value is only $2 \% $.
The numerical results are presented in figure \ref{cebronfig3}. In
order to compare them with the inviscid analytical prediction
(\ref{eq:sigma_c}), a viscous damping term $-K \sqrt{E}$ is added
to the expression of the inviscid growth rate, similarly to the
classical expression (\ref{eq:sigmasimple}). Excellent agreement is
then found for $a \leq c \leq b$, again validating our approach.
However, a slightly different constant $K=2.5$ (instead of $2.62$ in
(\ref{eq:sigma_c}), valid in the limit of small $\varepsilon$)
allows a better fit of our data. For strong deformations, this constant 
clearly depends on the considered ellipticity; for instance, a similar 
study performed at $\varepsilon=0.42$ implies $K=2.78$. Nevertheless, one 
can notice that $K$ always remains about the same order of magnitude. 
One can also notice that because of the scattering in the numerical results
presented in figure \ref{cebronfig2bis} (a), such a constant $K=2.5$
works equally well. Hence in the following, we systematically use
$K=2.5$.

In addition to the verification of the theoretical law of
\cite{Gledzer92}, we are now in the position of exploring the range
outside $a \leq c \leq b$, where other modes may grow that are not
necessarily linear in space variables. As shown in figure
\ref{cebronfig3}, different modes of the elliptical instability,
characterized by their main frequency of oscillation, appear
depending on the ratio c/b. In this view, the variation of the
oblateness is similar to the variation of the aspect ratio in the
case of an elliptically deformed cylinder \cite[][]{Eloy_2003}.
Since the elliptical instability comes from the parametric resonance
of two inertial waves of azimuthal wave number $m$ and $m+2$ with the
underlying strain field, the corresponding mode is written $(m,m+2)$
and is related to a pulsation of frequency $\omega_{mode}=m+1$. For
instance, the spin-over mode corresponds to the stationary mode
$(-1,1)$ with half a wavelength along the axis of rotation.
According to figure \ref{cebronfig3} (b), the mode (1,3) can be
observed when $c<b$ and the mode (0,2) when $c>a$. For even larger
aspect ratio, the mode (-1,1) with a larger wavenumber than the
spin-over takes place. Note however that because of the geometrical
confinement, no stationary mode can be excited for $c<b$.

\subsection{Influence of a background rotation}\label{sec:background}

In geo- and astrophysics, the tidal deformation of a given body
(planet, moon or star) is also in rotation because of the orbital
motion of the companion body. The influence of this background
rotation on the development of the elliptical instability has been
studied theoretically, using short-wavelength analysis
\cite[][]{Craik_1989,Leblanc_1997,LeDizes_2000} or normal mode
analysis \cite[][]{Gledzer92,Kerswell_1994}, and also numerically
for specific vortices such as Stuart vortices
\cite[][]{Leblanc_1998,Potylitsin_1999} or Taylor-Green vortices
\cite[][]{Sipp_1999}. It has been studied experimentally in
deformed cylinders \cite[][]{Vladimirov_1983,LeBars07} and ellipsoids 
\cite[][]{Boubnov_1978,LeBars09}. All these works show that the
background rotation has a stabilizing effect on cyclones and a
destabilizing effect on anticyclones, except when the background
rotation almost compensates for the flow rotation, in which case the
elliptical instability disappears.

\begin{figure}                   % Chaque figure doit avoir pour nom nomfig1.eps,
  \begin{center}
    \begin{tabular}{ccc}
      \setlength{\epsfysize}{6.5cm}
      \subfigure[]{\epsfbox{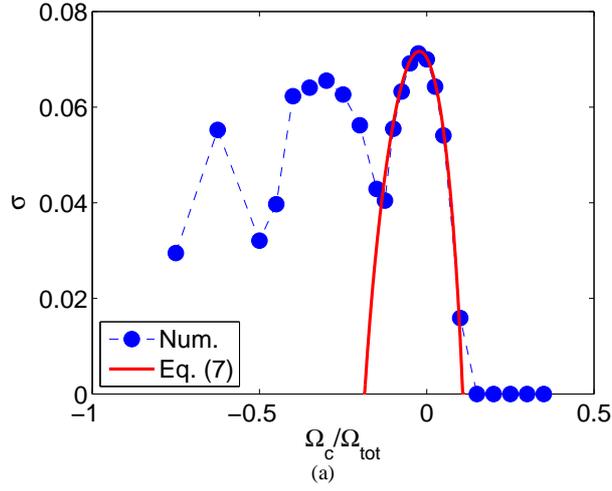}} \\
      \setlength{\epsfysize}{6.5cm}
      \subfigure[]{\epsfbox{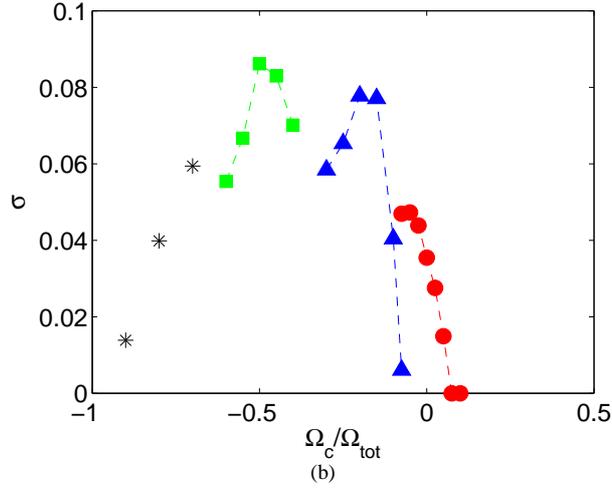}}
%       \setlength{\epsfysize}{4.5cm}
%       \subfigure[]{\epsfbox{A_c.eps}}
    \end{tabular}
    \caption{\it{Evolution of the growth rate in the presence of a background rotation for a fixed ellipticity $ \varepsilon=0.317 $ and a fixed
    value of the total Ekman number ${E}_{tot}=\frac{\nu }{\Omega_{tot}\ R_{eq}^2}=10^{-3}$, where
    $\Omega_{tot}$ takes into account the background rotation and the fluid rotation, i.e. $\Omega_{tot}=\Omega_c+\Omega$. (a) $c=(a+b)/2$, i.e.
$c/b=0.86$: the spin-over mode is then excited in the absence of
    background rotation. Good agreement is found around this value with the analytical solution (7). Further decreasing $\Omega_c$, other modes with smaller 
wavelength along the rotation axis appear. (b) $c/b=0.65$: the $(1,3)$ mode is then excited in the absence of
    background rotation. Other modes can be excited: $\bullet$ represent the $(1,3)$ mode, $\blacktriangle$ the spin-over mode, $\blacksquare$ 
represent the $(-1,1)$ mode with one wavelength along the axis of rotation, and $*$ are other modes.}}
    \label{cebronfig7}             % Pensez � mettre le nom du premier auteur � la place de Nom
  \end{center}
\end{figure}

The theoretical expression of the growth rate for perturbations that
are linear in space variables (i.e. corresponding to the classical
spin-over mode) can be readily obtained by following the same method
as \cite{Gledzer92}, but taking into account an additional Coriolis
force coming from the background rotation at a given angular
velocity $\Omega_c=\Omega_c^*\Omega$:
\begin{eqnarray}
  \sigma=\sqrt{\frac{(b^2-c^2+2\ \Omega_c^*\ a\ b )(c^2-a^2 - 2\ \Omega_c^*\ a\ b )}{(b^2+c^2)(a^2+c^2)}}.\label{eq:sigma_c2}
\end{eqnarray}
Actually, this is a particular case of the more general stability analysis detailled in 
\cite{Cebron_2010b} for a precessing triaxial ellipsoid. The existence range of this mode is now between $c=b \sqrt{1+2
\Omega_c^* a/b}$ and $c=a \sqrt{1+2\ \Omega_c^*\ b/a}$ and the maximal
theoretical growth rate is now exactly obtained for 
$\frac{c^2}{a\ b}=\Omega_c^*+\sqrt{1+\Omega_c^*  \left(a/b+b/a+\Omega_c^* \right)}$. In our numerical model, the addition of such a Coriolis 
force is straightforward and numerical results are presented in figure
\ref{cebronfig7} (a). Again, they compare well with the theoretical
prediction, providing that the viscous damping term, $-2.5 \sqrt{E}$,
found in section \ref{influence_length} is added to the inviscid
expression (\ref{eq:sigma_c2}). Other modes are selected by
the Coriolis force outside of the spin-over mode resonance band, as already
observed experimentally by \cite{LeBars09}.

Finally, we can investigate how the oblateness and the background
rotation interact with each others. Considering for instance an
oblate ellipsoid with a ratio $c/b$ such as the mode $(1,3)$ is
selected in absence of background rotation, the spin-over mode is
recovered when decreasing $\Omega_c$, as shown in figure
\ref{cebronfig7} (b). More generally, we find that the $(-1,1)$ mode
is the most unstable one in the anticyclonic domain up to a value $\Omega_c/\Omega_{tot} \lesssim
-1$ where the flow restabilises, in agreement with the conclusions
of \cite{LeBars09}.

\subsection{Influence of the obliquity of the ellipsoid}\label{obliquity}

In planetary cores, tidal deformations are aligned with the orbital
plane rather than with the equatorial plane, which means that they
are not orthogonal to the rotation axis. Previous works do not take
into account this phenomenon, even though the obliquity can be significant, e.g. ${23}^\circ 26'$ for the Earth. In this section, we thus 
investigate the effect of obliquity for the first time and consider that, as a first approximation, the shape of the 
body remains a triaxial ellipsoid.

\begin{figure}                   % Chaque figure doit avoir pour nom nomfig1.eps,
  \begin{center}
    \begin{tabular}{ccc}
      \setlength{\epsfysize}{7.5cm}
      \subfigure[]{\epsfbox{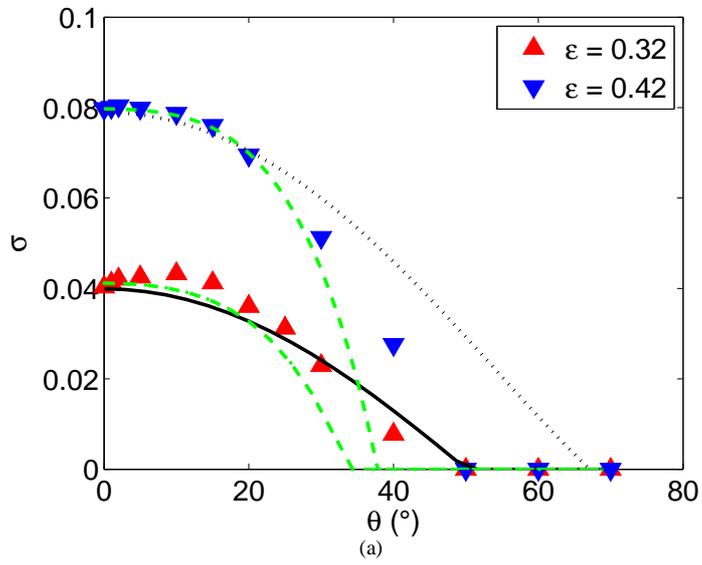}}
    \end{tabular}
    \caption{\it{Evolution of the growth rate with the obliquity $\theta$ for $E=1/600$ (fixed angular rate) and two values of the ellipticity $\varepsilon=0.317$ and $\varepsilon=0.42$.
    The dashed dotted line and the dashed line correspond to expression (\ref{eq:sigma_c}), taking into account the values of apparent axes seen from the equatorial plane,
    whereas the dotted line and the continuous line correspond to expression
    (\ref{eq:sigmasimple}), simply considering the apparent value of the ellipticity seen from the equatorial plane. The coefficient K  for viscous corrections is determined at $\theta=0^{\circ}$ and then kept constant. }}
    \label{cebronfig9}             % Pensez � mettre le nom du premier auteur � la place de Nom
  \end{center}
\end{figure}

In the numerical model, the rotation axis is tilted and oriented along the
unit vector $$\mathbf{k_c}=(\cos(\phi)\ \sin(\theta),\ \sin(\phi)\
\sin(\theta),\ \cos(\theta)),$$ where $\phi$ is its azimuth angle and
$\theta$ its colatitude angle in spherical coordinates (fig. \ref{cebronfig0}). In assuming
for example that the rotation axis is tilted in the $(Oxz)$ plane (i.e.
$\phi=0$), we can study how the development of the elliptical
instability changes depending on the obliquity $\theta$. For
$\theta=0$, one recovers the usual configuration with
$\varepsilon=(b^2-a^2)/(b^2+a^2)$ and an aspect ratio $c/b$. For
$\theta=\pi /2$, one recovers results from section
\ref{influence_length} in exchanging $a$ and $c$, i.e.
$\varepsilon=|b^2-c^2|/(b^2+c^2)$ and the aspect ratio is now
$a/b$. In between, streamlines in planes perpendicular to the
rotation axis are also elliptical, but their centers are not located along
the rotation axis anymore, except in the equatorial plane $z=0$. Besides
the effective ellipticity measured in each plane now depends on
$z$, the maximum being reached in the equatorial plane $z=0$. Seen
from this equatorial plane, the apparent length of the ellipsoid
along the rotation axis is $\tilde{c}={(\frac{\sin^2
\theta}{a^2}+\frac{\cos^2 \theta}{c^2})}^{-\frac{1}{2}}$, whereas the
great and small axes of the elliptical streamlines are respectively
$\tilde{b}=b$ and $\tilde{a}={(\frac{\cos^2 \theta}{a^2}+\frac{\sin^2
\theta}{c^2})}^{-\frac{1}{2}}$. One can then estimate the growth
rate of the instability using formulas
(\ref{eq:sigmasimple}) or (\ref{eq:sigma_c}) with these apparent
lengths. Numerical results for the spin-over mode compared with these approximations are
presented in figure \ref{cebronfig9}. Results agree well for
obliquity up to $\theta \sim {20}^{\circ}$. Then, the numerical
growth rate significantly differs, probably because the geometry is
 very far from an apparent ellipsoid in rotation around one of
its principal axes. It remains however between the two expressions
proposed. Note that for the Ekman number studied here, we do not
see the tidal instability reappearing around $\theta \sim
{90}^{\circ}$, which will however be the case at smaller Ekman
number: as seen in section \ref{influence_length},  in this case other modes could
appear because of the modified oblateness.

\section{Scaling laws for global quantities of geophysical interest}\label{quantif}

\subsection{Scaling law for the amplitude of the flow driven by the instability}\label{amplitude_sec}

\begin{figure}                   % Chaque figure doit avoir pour nom nomfig1.eps,
  \begin{center}
    \begin{tabular}{ccc}
      \setlength{\epsfysize}{7.5cm}
      \subfigure[]{\epsfbox{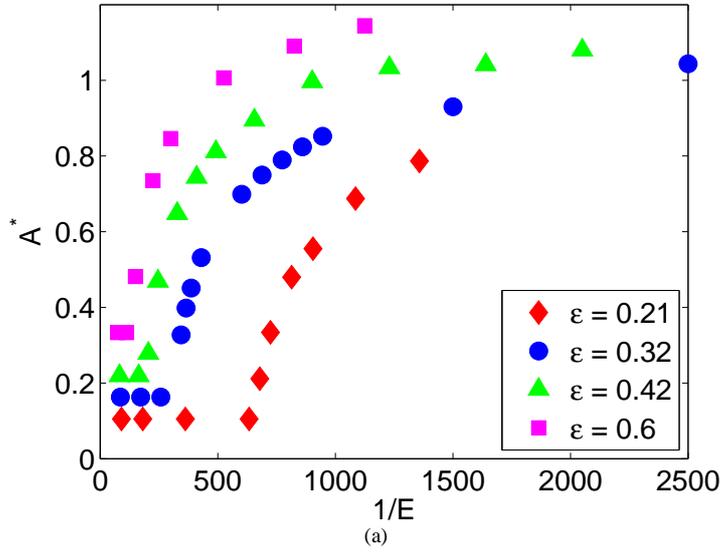}} \\
      \setlength{\epsfysize}{7.5cm}
      \subfigure[]{\epsfbox{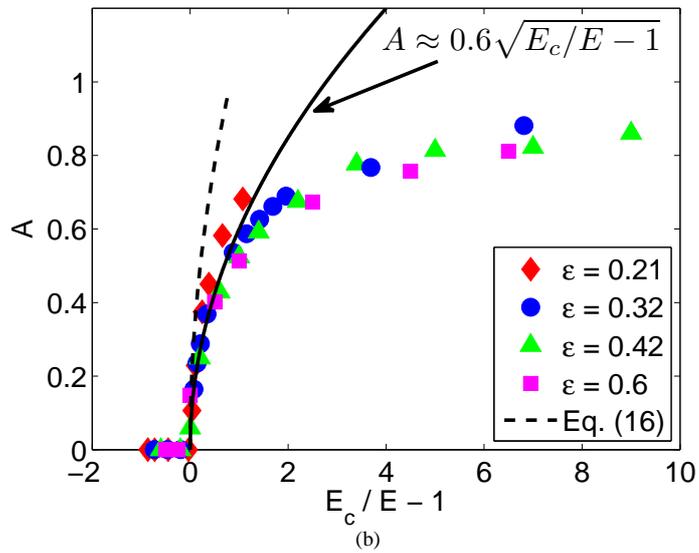}}
    \end{tabular}
    \caption{\it{Variation of the saturation amplitude of the flow driven by the elliptical instability depending on the Ekman number, for various values of the ellipticity.
    (a) Maximum value of the difference between the actual velocity and the theoretical base flow (\ref{baseflow}). (b) Maximum amplitude of the tidal instability,
    corresponding to the previous values corrected to take into account
    recirculation cells: all curves then superimpose when computed as a function
    of $E_c/E-1$.}}
    \label{cebronfig4}             % Pensez � mettre le nom du premier auteur � la place de Nom
  \end{center}
\end{figure}

The amplitude of the flow driven by the instability at saturation remains up to now an open
question, mainly because it is determined by strong non-linear
interactions. It is however an important quantity for geo- and
astrophysical applications, since it allows us to evaluate the
influence of elliptical aspects compared to the other relevant
ingredients of planetary dynamics, such as convection. 
In order to study the amplitude of the flow driven by the instability at saturation, we define 
the amplitude $A^*$ by the maximum value over the volume $V$:
\begin{eqnarray}
 A^*=\displaystyle \max_{V} ||\mathbf{u} - \mathbf{u_b} ||
\end{eqnarray}
where $\mathbf{u}$ is the dimensionless velocity field and
$\mathbf{u_b}$ is the theoretical base flow defined in section
\ref{influence_length}. The evolution of $A^*$ as a function of the
Ekman number is shown in figure \ref{cebronfig4} (a) for various
ellipticities: it can be seen that $A^*$ is not zero below the
threshold of the elliptical instability because of the differences around the outer bound between the theoretical base flow and the flow numerically obtained
(see figure \ref{cebronfig6}). Actually, the difference is maximal on the boundary, along the smallest equatorial axis where the theoretical velocity is maximal. There, the corresponding amplitude $\lambda_{(\varepsilon)}$ can be calculated theoretically : 
\begin{eqnarray}
\lambda_{(\varepsilon)}=\frac{b}{\frac{a+b}{2}}-1=\frac{2}{1+\sqrt{\frac{1-\varepsilon}{1+\varepsilon}}}-1
\end{eqnarray}
which only depends on $\varepsilon$. One then defines the amplitude of the flow driven by the 
instability by $A=A^*-\lambda_{(\varepsilon)}$, which is equal to $0$ below the threshold (fig. \ref{cebronfig4} (b)).

% In order to
% study the amplitude of the instability at saturation, we first
% define the amplitude $ A^*$ by
% \begin{eqnarray}
%  A^*=\displaystyle \max_{V} ||\mathbf{u} - \mathbf{u_b}||,
% \end{eqnarray}
% where $\mathbf{u}$ is the dimensionless velocity field and
% $\mathbf{u_b}$ is the base flow defined in section
% \ref{influence_length}. The evolution of $A^*$ as a function of the
% Reynolds number is shown in figure \ref{cebronfig4} (a) for various
% ellipticities. Due to the recirculations around the outer bound
% (see figure \ref{cebronfig6}), this amplitude is not zero below the
% threshold of the elliptical instability. Fortunately, numerical
% results show that this non-zero amplitude $\lambda_{(\varepsilon)}$ is
% only dependant on $\varepsilon$ and is found to be
% $\lambda_{(\varepsilon)}=0.358 \varepsilon +0.784 \varepsilon^2 $. Then, with
% this expression, the amplitude $A$ of the instability at saturation
% can be defined by
% \begin{eqnarray}
%  A=\displaystyle \max_{V} ||\mathbf{u} - \mathbf{u_b} || -\lambda_{(\varepsilon)}=A^*-\lambda_{(\varepsilon)}
% \end{eqnarray}

Far from threshold, a secondary instability appears, which induces a secondary dynamics superimposed on the primary state : 
e.g. typically, for $\varepsilon=0.317$, $E=1/1500$, $c=(a+b)/2$, the spinover mode is no more stationary, and the flow 
is slightly oscillating around the spinover mean flow, at a pulsation $\omega_{sec} \approx 1.4$. Note that this is in 
agreement with \cite{kerswell_2002} which predicts that the primary elliptical instability should saturate and be stable only in a 
small strain window $\varepsilon-\varepsilon_c=O(E)$. Thus, the amplitude $A$ has to be averaged in time to smooth
out the small scale fluctuations. Since those small-scale
fluctuations take place on a typical time scale comparable to one
revolution $2 \pi/\Omega$, whereas the characteristic time for the
tidal instability is the inverse of the growth rate, i.e. $(\Omega\
\varepsilon/2)^{-1}$, the averaging is performed about a typical time
$\Omega^{-1} \sqrt{4 \pi/\varepsilon}$ corresponding to the geometrical
mean of these two extreme values.

With this definition, figure \ref{cebronfig4} (b) shows that all results collapse
on the same generic law above the threshold providing that we use the variable $E_c/E-1$. Far 
from threshold, the amplitude seems to saturate around 1, which means that the velocities generated by the
tidal instability are comparable to the imposed boundary rotation.

One can notice that near
the threshold, a square root $A \approx 0.6\ \sqrt{E_c/E-1} $ fits the
results, which is in agreement with a pitchfork bifurcation. Actually, this scaling law can be obtained 
analytically, starting from the simple model used in \cite{lacaze_2004} to describe the viscous non-linear evolution of the spinover mode. This model, 
first introduced by \cite{Hough_1895} and \cite{Poincare} for an inviscid solid-body rotation in a spheroid, reduces to:
\begin{eqnarray}
\dot{\omega_x}=- \alpha\ (1+\omega_z)\ \omega_y-\nu_{so}\ \omega_x
\end{eqnarray}
\begin{eqnarray}
\dot{\omega_y}=- \beta\ (1+\omega_z)\ \omega_x-\nu_{so}\ \omega_y
\end{eqnarray}
\begin{eqnarray}
\dot{\omega_z}=\varepsilon\ \omega_x\ \omega_y-\nu_{ec}\ \omega_z+\nu_{nl}\ (\omega_x^2+\omega_y^2)
\end{eqnarray}
where $\boldsymbol{\omega}$ is the rotation vector of the spinover mode, $\alpha=\frac{\varepsilon}{2-\varepsilon}$ and $\beta=\frac{\varepsilon}{2+\varepsilon}$. 
The damping terms are given by theory (no adjustable parameter): $\nu_{so}=2.62\ \sqrt{E}$ is the linear viscous damping rate of the spinover mode (first calculated by \cite{Greenspan}), $\nu_{ec}=2.85\ \sqrt{E}$ is the linear viscous damping of axial rotation and $\nu_{nl}=1.42\ \sqrt{E}$ is the viscous boundary layer effect on the non-linear interaction of the spinover mode with itself (see \cite{Greenspan}). Even if this model does not take into account all the viscous terms of $O(\sqrt{E})$ or the non-linear corrections induced by the internal shear layers, it satisfyingly agrees with experiments, regarding the growth rate as well as the non-linear saturation \cite[][]{lacaze_2004}. Thus, this model can be used with confidence to describe the viscous non-linear evolution of the spinover mode. 

After little algebra, we obtain a non-trivial stationary state for $\varepsilon > 2 \nu_{so}$, given by:

\begin{eqnarray}
\omega_x=\pm\ \sqrt{\frac{\nu_{ec}\ [\sqrt{\alpha \beta}-\nu_{so}] } {\beta\ \varepsilon-\nu_{nl}\ [\sqrt{\alpha \beta} +\beta^2 / \sqrt{\alpha \beta}] }} 
\approx \pm\ \sqrt{\frac{\nu_{ec}\ [\varepsilon-2\ \nu_{so}] } {\varepsilon^2-2\ \nu_{nl}\ \varepsilon }} 
\end{eqnarray}
\begin{eqnarray}
\omega_y=\mp\ \sqrt{\frac{\nu_{ec}\ [\sqrt{\alpha \beta}-\nu_{so}] } {\alpha\ \varepsilon-\nu_{nl}\ [\sqrt{\alpha \beta} +\alpha^2 / \sqrt{\alpha \beta}] }}
\approx \mp\ \sqrt{\frac{\nu_{ec}\ [\varepsilon-2\ \nu_{so}] } {\varepsilon^2-2\ \nu_{nl}\ \varepsilon }}  \approx \mp\ \omega_x
\end{eqnarray}
\begin{eqnarray}
\omega_z=\frac{\nu_{so}}{\varepsilon}\ \sqrt{4-\varepsilon^2}-1 \approx \frac{2\ \nu_{so}}{\varepsilon}-1
\end{eqnarray}
where approximations are done assuming $\varepsilon \ll 1$. Now, the amplitude $A$ corresponds to the norm of the flow $\boldsymbol{\omega} \times \boldsymbol{r}$ driven by the spinover mode. Then, near the threshold ($\varepsilon \approx 2 \cdot 2.62 \sqrt{E_c}$ and $E_c/E-1 \ll 1$), we obtain: 
\begin{eqnarray}
A \approx \sqrt{\frac{\nu_{ec}}{2\ (\nu_{so}-\nu_{nl})} \left( \frac{E_c}{E}-1 \right) }  \approx 1.1\ \sqrt{\frac{E_c}{E}-1}
\end{eqnarray}
which is in good agreement with the numerical fit.

\subsection{Scaling law for the viscous dissipation by the instability}

As explained for instance in \cite{Rieutord} or in \cite{LeBars09},
the energy dissipated by tides is a primordial quantity which
directly influences the orbital evolution and rotational history of
a binary system during its synchronization. It is however poorly
known \cite[e.g.][]{williams_2000,touma_1994}. In most traditional
models, fluid dissipation in the planetary core is supposed to be
negligible. However this may not be the case when the elliptical
instability is excited at a rather large amplitude, inducing
important shears between the bulk of the fluid and the boundary. Our
purpose here is to systematically quantify the variation of this
dissipation with the ellipticity and the Ekman number.

The dissipated power balance of the incompressible flow in the ellipsoid is given by:
\begin{eqnarray}
\iiint_V \frac{\partial }{\partial t} \left( \rho\ \frac{\tilde{u}^2}{2} \right)\ \D \tau = \iint_S \left( \boldsymbol{\bar{\bar{\sigma}}_v} \cdot \mathbf{n}  \right) \cdot \mathbf{\tilde{u}}\ \D s 
- \iiint_V \boldsymbol{\bar{\bar{\sigma}}_v} : \boldsymbol{\nabla} \mathbf{\tilde{u}}\ \D \tau \label{eq:balance}
\end{eqnarray}
where $\boldsymbol{\bar{\bar{\sigma}}_v}=\eta \left( \boldsymbol{\nabla} \mathbf{\tilde{u}} + \mathstrut^t \boldsymbol{\nabla} \mathbf{\tilde{u}} \right)$ is the viscous stress tensor of the newtonian fluid, $S$ the surface of the ellipsoid, $\mathbf{\tilde{u}} = \Omega\ R_{eq}\ \mathbf{u}$ is the
dimensionalized velocity field, $\rho$ the volumic mass and $\eta$ the dynamic viscosity of
the fluid. In this section, we focus on the stationary spinover mode and then the two terms on the right side of equation (\ref{eq:balance}) balance each others: the first one allows to maintain the rotation of the fluid by the no-slip conditions on the boundary while the second one is the volume dissipation $P$ of the fluid. In the following, we use as a power scale the dissipated
power during the spin-down stage of the equivalent sphere of radius
$R_{eq}$ and moment of inertia $I_{\Delta}$ ($I_{\Delta}=2/5\ M R^2$ for an homogeneous sphere of mass $M$ and radius $R$), i.e. the kinetic energy
of rotation $E_c=\frac{1}{2}\ I_{\Delta}\ \Omega^2$ divided by the
Ekman time $t_{E}$. Note that most of the dissipated power comes
naturally from the boundary layers: its determination is thus a
challenging task from a numerical point of view, and extra care must
be taken regarding the convergence of the simulations, as shown for
instance in figure \ref{cebronfig1} (b).

Below the threshold, the dissipated power is not zero because of the recirculation patterns 
of our base flow (see figure \ref{cebronfig6}). This dissipation is due to the flow driven by the ellipticity, which scales 
as $\varepsilon\ \Omega R$ for small ellipticities. Then, the dimensionalized dissipated power below the threshold simply scales as 
\begin{eqnarray}
\iiint_V \boldsymbol{\bar{\bar{\sigma}}_v} : \boldsymbol{\nabla} \mathbf{\tilde{u}}\ \D \tau \sim \eta\ \varepsilon^2\ \Omega^2\ R^3 
\end{eqnarray}
which is confirmed by the dissipated power measured in our numerical simulations below the threshold. 

Far from threshold, the model proposed by \cite{LeBars09} considers
that the spin-over mode simply corresponds to a supplementary solid
body rotation $ \bf {\Omega}_{SO}$ outside the outer viscous
boundary layer of thickness $ h= \xi \sqrt{\nu / \Omega_{SO}} $, $\xi$
being a constant of order $1$. 
Then, the dissipation is only located in this
boundary layer, where the fluid rotation has to match the imposed
velocity conditions at the outer boundary. With this simple model,
the torque of the container on the fluid is $\mathbf{C^{m/c}}=-2\ M\
\nu\ \frac{R}{h}\ \mathbf{{\Omega}_{SO}}$, and the power dissipated by
the system is $P= -2\ M\ \nu\ \frac{R}{h}\ {\Omega_{SO}}^2$, which can
also be written $P= -\frac{2}{\xi}\ M\ R\ \sqrt{\nu}\ \Omega^{5/2}\ A^{5/2}$,
where $A$ is the dimensionless spin-over mode amplitude studied in
the previous section. The dimensionless power given by this model 
can finally be stated as
\begin{eqnarray}
 {P}_{dissip}= \frac{|P|}{\frac{1}{2}\ I_{\Delta}\ \Omega^3\ \sqrt{E}} = \frac{10}{\xi}\
 A^{5/2},\label{scalingpower}
\end{eqnarray}
where, according to the previous section, $A$ is about $1$ in the
small Ekman number limit studied here.

\begin{figure}
  \begin{center}
    \epsfysize=10.0cm
    \leavevmode
    \epsfbox{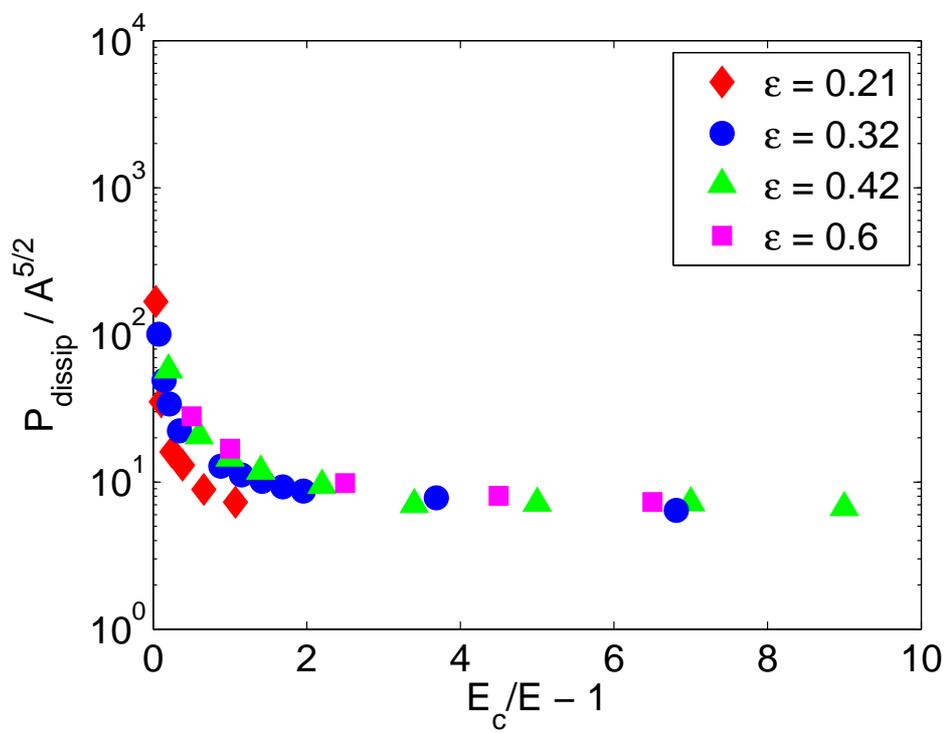}
    \caption{\it{Viscous dissipation by the tidal instability, as a function of the distance from threshold.
    The results collapse on a generic law and seem to converge towards a saturation value far from threshold.}}
    \label{cebronfig5}
  \end{center}
\end{figure}

The evolution of $P_{dissip}$ measured in our numerical model is
shown in figure \ref{cebronfig5}. It confirms the
behavior predicted by (\ref{scalingpower}) with a saturation far from threshold 
and $\xi \approx 1$. This behavior is
especially interesting for planetary applications, as will be
studied in the following section.

\section{Orders of magnitude for planetary applications}\label{discussion}

From a geophysical point of view, the results from the previous
sections can be used to derive the orders of magnitude involved in
planetary cores. Regarding the jovian moon Io, which is clearly
unstable to the tidal instability and where the strong tidal
dissipation is a topical question \cite[][]{lainey_2009}, our
numerical study confirms the first trends given by \cite{LeBars09},
with however a numerically determined scaling factor of the order $1/\xi \approx 1 $ in the
power dissipation estimates. Moreover, the most important result comes
from the oblateness effects. Indeed, according to the data given in
\cite{KerswellMalkus}, $c/b \sim 0.995$ and $a/b > 0.999$, which
means that the excited mode of the tidal instability in Io cannot be
the spin-over mode, but has to be oscillatory. Note that this result 
is not modified when the background rotation is taken into account.

A similar analysis can be done for the early Earth, with a Moon two
times closer than today. In this case, the length of the day was
around 10 hours according to \cite{touma_1994} and the lunar tides
were around eight times stronger, whereas the solar tides were about
the same. Then, the actual tidal amplitude of $50$ cm allows to
estimate the ellipticity of the early Earth: $\varepsilon \sim
10^{-6}$. In considering a similar but totally molten core, we
estimate $E_c/E-1 \sim 70$, which means that the early Earth's
core was clearly unstable to tidal instability, and that the tidal
instability was then at saturation. Moreover, with an orbital period
for the Moon $(1/2)^{3/2}$ shorter than the actual moon (Kepler's
Third Law), the dissipated power given by the model was around $P
\sim 10^{18} \ W$. This estimation seems huge in comparison with the 
present dissipation by tidal friction ($\sim 3.75 \cdot 10^{12}\ W$ according to \cite{Munk}) but actually, it
simply suggests that the Earth-Moon system was then in rapid
evolution. Once again, the possible mode of the tidal instability
was not the spin-over mode neither any stationary $(-1,1)$ mode,
considering that the oblateness of the rapidly rotating early Earth
was larger than the actual oblateness. In fact, we expect stationary
modes to be only marginally excited in geophysical systems, limited to 
very peculiar configuration regarding the oblateness and the rotation of the tidal bulge.

Note that computations similar to the ones presented here for
an ellipsoidal geometry have been performed for an ellipsoidal shell
in order to study the influence of a solid inner planetary core. Similar
conclusions are then found regarding the physics of the tidal
instability. Only a slightly larger dissipation term due to the
presence of an inner viscous boundary layer has to be taken into
account. 

\section{Conclusion}

This paper presents the first systematic numerical
study of the tidal instability in a rotating ellipsoid. The
numerical approach is a powerful tool to complete the 
knowledge derived from previous theoretical and experimental
studies. The effects of oblateness, background rotation and
obliquity, which lead to the selection of various
instability modes, have been studied. We have also defined scaling laws regarding the
amplitude of the flow driven by the instability and its viscous dissipation, which
confirm the primordial role played by tidal effects at a planetary
scale.

Another fundamental geophysical issue regarding the elliptical
instability is whether or not it can drive a planetary dynamo. The
results presented herein show that the numerical approach is able to
faithfully capture the physics of the hydrodynamic elliptical
instability. Thus, the present work provides a solid basis to study
the full MHD problem encountered in planetary core flows undergoing
tidal instability. These flows can only be studied numerically due to
their inherent complexity. Therefore, a next step will be to introduce
the effects of a magnetic field in our model.

\newpage

% \begin{figure}
%   \begin{center}
%     \epsfysize=5.0cm
%     \leavevmode
%     \epsfbox{sig_sphere.eps}
%     \caption{Validation of the numerical simulations : evolution of the growth rate and comparison with the linear theory ($ 0 \leq Re \leq 1720$, $0.05 \leq \varepsilon \leq 0.9$, $c=\frac{a+b}{2} $).}
%     \label{cebronfig2bis}
%   \end{center}
% \end{figure}

\begin{thebibliography}{00}

\bibitem[{Aldridge et~al.(1997)Aldridge, Seyed-Mahmoud, Henderson, and van Wijngaarden}]{Aldridge}
Aldridge, K., Seyed-Mahmoud, B., Henderson, G., van~Wijngaarden, W.,
1997. Elliptical instability of the {Earth}'s fluid core. Phys.
Earth Planet. Int. 103, 365--74.

\bibitem[{Bayly(1986)}]{Bayly_1986}
Bayly, B.~J., 1986. Three-dimensional instability of elliptical flow. Phys. Rev. Lett. 57, 2160 - 2163

\bibitem[{Benton and Clark(1974)}]{benton}
Benton, E.R., Clark, A., 1974. Spin-Up. Annu. Rev. Fluid. Mech. 6, 257--280.

\bibitem[{Boubnov(1978)}]{Boubnov_1978}
Boubnov, B.~M., 1978. Effect of Coriolis force field on the motion of a fluid inside an ellipsoidal cavity. Izv. Atmos. Ocean. Phys, 14, 501-504.

\bibitem[{Cadot et~al.(2009)Cadot, Douady, Couder}]{Cadot}
Cadot, O., Douady, S., Couder, Y., 1995. Characterization of the low pressure filaments in three-dimensional turbulent shear flow. Phys. Fluids 7, 630--646.

\bibitem[{C\'ebron et~al.(2010a)C\'ebron, Maubert, Le~Bars}]{Cebron_2010}
C\'ebron, D., Maubert, P., Le~Bars, M., 2010. Tidal instability in a rotating and differentially heated ellipsoidal shell. Geophys.\ J.\ Int. In revision.

\bibitem[{C\'ebron et~al.(2010b)C\'ebron, Le~Bars, Meunier}]{Cebron_2010b}
C\'ebron, D., Le~Bars, M., Meunier, P., 2010. Tilt-over mode in a precessing triaxial ellipsoid. Phys. Fluids. In revision.


\bibitem[{Craik(1989)}]{Craik_1989}
Craik, A.~D.~D., 1989. The stability of unbounded two- and three-dimensional flows subject to body forces: some exact solutions. J. Fluid Mech. 198, 275-292.

\bibitem[{Chan et~al.(2010)Chan, Zhang, Liao}]{Chan_2010}
Chan, K.~H, Zhang, K., Liao, X., 2010. An EBE finite element method for simulating nonlinear flows in rotating spheroidal cavities. International Journal for Numerical Methods in Fluids
Volume 63 Issue 3, Pages 395 - 414


\bibitem[{Eloy et~al.(2000)Eloy, Le~Gal, and Le~Diz\`es}]{Eloy_2000}
Eloy, E., Le~Gal, P., Le~Diz\`es, S., 2000. Experimental study of the multipolar vortex instability. Phys. Rev. Lett. 85, 145--166.

\bibitem[{Eloy et~al.(2003)Eloy, Le~Gal, and Le~Diz\`es}]{Eloy_2003}
Eloy, E., Le~Gal, P., Le~Diz\`es, S., 2003. Elliptic and triangular instabilities in rotating cylinders. J. Fluid Mech. 476, 357--388.

\bibitem[{Greenspan(1968)}]{Greenspan}
Greenspan, H.~P., 1968. The Theory of Rotating Fluids, Cambridge University Press, Cambridge.

\bibitem[{Herreman et~al.(2009)Herreman, Le~Bars, Le~Gal}]{herreman}
Herreman, W., Le~Bars, M., Le~Gal, P., 2009. On the effects of an imposed magnetic field on the elliptical instability in rotating spheroids. Phys. Fluids 21, 046602.

% \bibitem[{Herreman et~al.(2009)Herreman, Le~Bars, Le~Gal}]{herreman_2010}
% Herreman, W., Le~Bars, M., Le~Gal, P., 2009. On the effects of an imposed magnetic field on the elliptical instability in rotating spheroids. Phys. Fluids 21, 046602.

\bibitem[{Gledzer and Ponomarev(1977)}]{Gledzer77}
Gledzer, E.~B., Ponomarev, V.~M., 1977. Finite dimensional approximation of the motions of incompressible fluid in an ellipsoidal cavity. Izv. Atmos. Ocean. Phys. 13, 565--569.

\bibitem[{Gledzer and Ponomarev(1992)}]{Gledzer92}
Gledzer, E.~B., Ponomarev, V.~M., 1992. Instability of bounded flows with elliptical streamlines. J. Fluid Mech. 240, 1--30.

\bibitem[{Gledzer et~al.(1975)Gledzer, Dolzhansky, Obukhov, Ponomarev}]{Gledzer_1975}
Gledzer, E.~B., Dolzhansky, F.~V., Obukhov, A.~M., Ponomarev, V.~M., 1975. An experimental and theoretical study of the stability of motion of a liquid in an elliptical cylinder. Izv. Atmos. Ocean. Phys. 11, 617--622.

\bibitem[{Haj-Hariri and Homsy(1997)}]{Haj}
Haj-Hariri, H., Homsy, G.~M., 1997. Three-dimensional instability of viscoelastic elliptic vortices. J. Fluid Mech. 353, 357--381.

\bibitem[{Hindmarsh et~al.(2005)Hindmarsh, Brown, Grant, Lee, Serban, Shumaker, Woodward}]{Hindmarsh_2005}
Hindmarsh, A.~C., Brown, P.~N., Grant, K.~E., Lee, S.~L., Serban, R., Shumaker, D.~E., Woodward, C.~S., 2005. SUNDIALS: Suite of Nonlinear and Differential/Algebraic Equation Solvers. ACM T. Math. Software, vol. 31,
p. 363.




\bibitem[{Hollerbach and Kerswell(1995)}]{Hollerbach_1995}
Hollerbach, R., Kerswell, R.~R., 1995. Oscillatory internal shear
layers in rotating and precessing flows. J. Fluid Mech. 298, 327--339.

\bibitem[{Hough(1895)}]{Hough_1895}
Hough, S.~S., 1895. The oscillations of a rotating ellipsoidal shell containing fluid. Phil. Trans. A 186, 469--506.


% \bibitem[{Kerswell(1993)}]{Kerswell93bis}
% Kerswell, R.~R., 1993. The instability of precessing flow. Geophys.
% Astrophys. Fluid Dyn. 72, 107--144.

% \bibitem[{Kerswell(1996)}]{Kerswell96}
% Kerswell, R.~R., 1996. Upper bounds on the energy dissipation in
% turbulent
%   precession. J. Fluid Mech. 321, 335--370.

\bibitem[{Kerswell(1994)}]{Kerswell_1994}
Kerswell, R.~R., 1994. Tidal excitation of hydromagnetic waves and their damping in the Earth. J. Fluid Mech. 274, 219--241.

\bibitem[{Kerswell(2002)}]{kerswell_2002}
Kerswell, R.~R., 2002. Elliptical instability. Annu. Rev. Fluid. Mech. 34, 83--113.

\bibitem[{Kerswell and Malkus(1998)}]{KerswellMalkus}
Kerswell, R.~R., Malkus, W. V.~R., 1998. Tidal instability as the
source for {Io}'s magnetic signature. Geophys. Res. Lett. 25, 603--6.

\bibitem[{Kelvin(1880)}]{Kelvin_1880}
Kelvin, L., 1880. Vibrations of a columnar vortex. Phil. Mag. 10,
   155--168.


\bibitem[{Kudlick(1966)}]{Kudlick}
Kudlick, M., 1966. On the transient motions in a contained rotating
fluid. PhD
  thesis, MIT.

\bibitem[{Lacaze et~al.(2006)Lacaze, Herreman, Le~Bars, Le~Diz\`es, and
  Le~Gal}]{Lacaze}
Lacaze, L., Herreman, W., Le~Bars, M., Le~Diz\`es, S., Le~Gal, P.,
2006. Magnetic field induced by elliptical instability in a rotating spheroid. Geophys. Astrophys. Fluid Dyn. 100, 299--317.

\bibitem[{Lacaze et~al.(2004)Lacaze, Le~Gal, and Le~Diz\`es}]{lacaze_2004}
Lacaze, L., Le~Gal, P., Le~Diz\`es, S., 2004. Elliptical instability
in a rotating spheroid. J. Fluid Mech. 505, 1--22.

\bibitem[{Lacaze et~al.(2005)Lacaze, Le~Gal, and Le~Diz\`es}]{lacaze_2005}
Lacaze, L., Le~Gal, P., Le~Diz\`es, S., 2005. Elliptical instability of the flow in a rotating shell . Phys. Earth. Planet. Int., 151, pp. 194-205

\bibitem[{Lacaze et~al.(2007)Lacaze, Ryan, and Le~Diz\`es}]{Lacaze_2007}
Lacaze, L., Ryan, K., Le~Diz\`es, S., 2007. Elliptic instability in a strained Batchelor vortex. J. Fluid Mech. 577, 341

\bibitem[{Lainey et~al.(2009)Lainey, Arlot, Karatekin and Van~Hoolst}]{lainey_2009}
Lainey, V., Arlot, J.~E., Karatekin, O., Van~Hoolst, T., 2009. Strong tidal dissipation in Io and Jupiter from astrometric observations. Nature 459, 957-959.

% \bibitem[{Le~Bars and Le~Diz\`es(2006)}]{LeBars06}
% Le~Bars, M., Le~Diz\`es, S., 2006. Thermo-elliptical instability in
% a rotating
%   cylindrical shell. J. Fluid Mech. 563, 189--198.

\bibitem[{Le~Bars et~al.(2010)Le~Bars, Lacaze, Le~Diz\`es, Le~Gal and Rieutord}]{LeBars09}
Le~Bars, M., Lacaze, M., Le~Diz\`es, S., Le~Gal, P., Rieutord, M., 2010. Tidal instability in stellar and planetary binary system. Phys. Earth. Planet. Int., 178, Issues 1-2, January 2010, Pages 48-55.

\bibitem[{Le~Bars et~al.(2007)Le~Bars, Le~Diz\`es, and Le~Gal}]{LeBars07}
Le~Bars, M., Le~Diz\`es, S., Le~Gal, P., 2007. Coriolis effects on
the elliptical instability in cylindrical and spherical rotating containers. J. Fluid Mech. 585, 323--342.


\bibitem[{Leblanc and Cambon(1997)}]{Leblanc_1997}
Leblanc, S., Cambon, C., 1997. On the three-dimensional instabilities of plane flows subjected to Coriolis force. Phys. Fluids 9, 1307.

\bibitem[{Leblanc and Cambon(1998)}]{Leblanc_1998}
Leblanc, S., Cambon, C., 1998. Effects of the Coriolis force on the stability of Stuart vortices. J. Fluid Mech. 356, 353-379.

\bibitem[{Le~Diz\`es(2000)}]{LeDizes_2000}
Le~Diz\`es, S., 2000. Three-dimensional instability of a multipolar
vortex in a rotating flow. Phys. Fluids 12, 2762--74.


\bibitem[{Le~Diz\`es and Laporte(2002)}]{LeDizes_2002}
Le~Diz\`es, S., Laporte, F., 2002. Theoretical predictions for the elliptic instability in a two-vortex flow. J. Fluid Mech. 471, 169--201.

\bibitem[{Leweke and Williamson(1998a)}]{Leweke98_a}
Leweke, T., Williamson, C.~H.~K., 1998. Three-dimensional instabilities in wake transition. Eur. J. Mech. B/Fluids 17, 571--586.


\bibitem[{Leweke and Williamson(1998b)}]{Leweke98_b}
Leweke, T., Williamson, C.~H.~K., 1998. Cooperative elliptic instability of a vortex pair. J. Fluid Mech. 360, 85--119.

\bibitem[{Lundgren and Mansour(1995)}]{lundgren}
Lundgren, T.~S., Mansour, N.~N., 1995. Transition to turbulence in an elliptic vortex. J. Fluid Mech. 307, 43--62.

\bibitem[{Malkus(1968)}]{malkus68}
Malkus, W. V.~R., 1968. Precession of the Earth as the cause of
geomagnetism.
  Science 160, 259--264.

\bibitem[{Malkus(1989)}]{malkus89}
Malkus, W. V.~R., 1989. An experimental study of global instabilities due to tidal (elliptical) distortion of a rotating elastic cylinder. Geophys. Astrophys. Fluid Dyn. 48, 123--134.

\bibitem[{Mason and Kerswell(1999)}]{mason}
Mason, D.~M., Kerswell, R.~R., 1999. Nonlinear evolution of the elliptical instability: an example of inertial wave breakdown. J. Fluid Mech. 396, 73--108.


\bibitem[{Meunier et~al.(2002)Meunier, Ehrenstein, Leweke and Rossi}]{Meunier_2002}
Meunier, P., Ehrenstein, U., Leweke, T., Rossi, M., 2002. A merging criterion for two-dimensional co-rotating vortices. Phys. Fluids 14, 2757--2766.


\bibitem[{McAlister et~al.(2005)McAlister, Ten~Brummelaar, Gies, Huang, Bagnuolo, Shure, Sturmann, Sturmann, Turner, Taylor, Berger, Baines, Grundstrom, Ogden, Ridgway and Van~Belle}]{mac_2005}
McAlister, H. A., Ten~Brummelaar, T. A., Gies, D. R., Huang, W. , Bagnuolo, W. G., Shure, M. A., Sturmann, J., Sturmann, L., Turner, N. H., Taylor, S. F., Berger, D. H., Baines, E. K., Grundstrom, E., Ogden, C., Ridgway, S. T., van Belle, G., 2005. First results from the CHARA array. I. An interferometric and spectroscopic study of the fast rotator $\alpha$ Leonis (Regulus). The Astrophysical journal, vol. 628 (1), no1, pp. 439-452.

\bibitem[{Moore and Saffman(1975)}]{Moore_1975}
Moore, D.~W., Saffman, P.~G., 1975. The instability of a straight vortex filament in a strain field. Proc. R. Soc. Lond. A 346, 413--425.

\bibitem[{Munk(1998)}]{Munk}
Munk, W., 1998. Abyssal recipes II: energetics of tidal and wind mixing. Deep Sea Res. Part I. Oceanographic Research, vol. 45, no12, pp. 1977-2010.

\bibitem[{Owen and Rogers(1989)}]{Owen_1989}
Owen, J.~M., Rogers, R.~H., 1989. Flow and heat transfer in rotating disc
systems (Vol. 1, Rotor-stator systems). Research Studies Press, Taunton, UK and John Wiley, NY.


\bibitem[{Pierrehumbert(1986)}]{pierrehumbert}
Pierrehumbert, R.~T., 1986. Universal short-wave instability of two-dimensional eddies in an inviscid fluid. Phys. Rev. Lett. 57, 2157 - 2159.

\bibitem[{Poincar\'e(1910)}]{Poincare}
Poincar\'e, R., 1910. Sur la pr\'ecession des corps d\'eformables. Bull. Astr. 27 (1910) 321.


\bibitem[{Potylitsin and Peltier(1999)}]{Potylitsin_1999}
Potylitsin, S., Peltier, W.~R., 1999. Three-dimensional destabilization of Stuart vortices: the influence of rotation and ellipticity. J. Fluid Mech. 387,205-226.

\bibitem[{Rieutord(2003)}]{Rieutord}
Rieutord, M., 2003. Evolution of rotation in binaries: physical
processes. Stellar Rotation, Proc. IAU Symp. 215, 394--403.

% \bibitem[{Rieutord et~al.(2001)Rieutord, Georgeot, and Valdettaro}]{Rieutord01}
% Rieutord, M., Georgeot, B., Valdettaro, L., 2001. Inertial waves in
% a rotating
%   spherical shell: attractors and asymptotic spectrum. J. Fluid Mech. 435,
%   103--144.

\bibitem[{Roy et~al.(2007)Roy, Schaeffer, Le~Diz\`es and Thompson}]{Roy}
Roy, M., Schaeffer, N., Le~Diz\`es, S., Thompson M.~C., 2007. Stability of a pair of co-rotating vortices with axial flow. Phys. Fluids 20, 094101.

\bibitem[{Seyed-Mahmoud et~al.(2004)Seyed-Mahmoud, Aldridge, and Henderson}]{Seyed_2004}
Seyed-Mahmoud, B., Aldridge, K.~D., Henderson, G., 2004. Elliptical instability in rotating spherical fluid shells: application to Earth's fluid core. Phys. Earth Planet. Int. 142, 257--282.

\bibitem[{Seyed-Mahmoud et~al.(2000)Seyed-Mahmoud, Henderson, and Aldridge}]{Seyed_2000}
Seyed-Mahmoud, B., Henderson, G., Aldridge, K.~D., 2000. A numerical model for elliptical instability of the Earth's fluid outer core. Phys. Earth Planet. Int. 117, 51--61.

\bibitem[{Shangli et~al.(2007)Shangli, Tohline, and Motl}]{Shangli}
Shangli, O., Henderson, J.~E., Motl, P.~M., 2007. Further evidence for an elliptical instability in rotating fluid bars and ellipsoidal stars.Astrophys. J. 665,1074


\bibitem[{Sipp and Jacquin(1998)}]{Sipp_1998}
Sipp, D., Jacquin, L., 1998. Elliptic instability in two-dimensional flattened Taylor--Green vortices. Phys. Fluids 10, 839.

\bibitem[{Sipp et~al.(1999)Shangli, Lauga, and Jacquin}]{Sipp_1999}
Sipp, D., Lauga, E., Jacquin, L., 1999. Vortices in rotating systems: Centrifugal, elliptic and hyperbolic type instabilities. Phys. Fluids, vol. 11, Issue 12, 3716.

% \bibitem[{Suess(1971)}]{Suess_1971}
% Suess, S.~T. 1971. Viscous flow in a deformable rotating container. J. Fluid Mech. 45, 189--201.

\bibitem[{Thess and Zikanov(2007)}]{thess}
Thess, A., Zikanov, O., 2007. Transition from two-dimensional to three-dimensional magnetohydrodynamic turbulence. J. Fluid Mech., 579, p. 383-412.

\bibitem[{Touma and Wisdom(1994)}]{touma_1994}
Touma, J., Wisdom, J., 1994. Evolution of the Earth-Moon system. The Astr. J., vol. 108, no. 5, p. 1943-1961.


% \bibitem[{Vanyo(1991)}]{vanyo}
% Vanyo, J.~P., 1991. A geodynamo powered by luni-solar precession.
% Geophys.
%   Astrophys. Fluid Dyn. 59, 209--234.
%
% \bibitem[{Vanyo and Likins(1972)}]{Vanyo72}
% Vanyo, J.~P., Likins, P.~W., 1972. Rigid-body approximations to
% turbulent motion in a liquid-filled, precessing spherical cavity. J.
% Appl. Mech., 39, 18--24.

\bibitem[{Vladimirov et~al.(1983)}]{Vladimirov_1983}
Vladimirov, V.~A., Tarasov, V.~F., Rybak, L.~Ia., 1983. The stability of the elliptically deformed rotation of an ideal incompressible fluid in a Coriolis force field. Izv. Atmos. Ocean. Phys, 14, 501-504.

\bibitem[{Waleffe(1990)}]{waleffe_1990}
Waleffe, F.~A., 1990. On the three-dimensional instability of strained vortices. Phys. Fluids 2, 76--80.

\bibitem[{Widnall et~al.(1974)Widnall, Bliss, and Tsai}]{Widnall_1974}
Widnall, S.~E., Bliss, D., Tsai, C.-Y, 1974. The instability of short waves on a vortex ring. J. Fluid Mech. 66, 35--47.

\bibitem[{Williams(2000)}]{williams_2000}          % Nom = nom du premier auteur
Williams, G.~E., 2000. Geological constraints on the Precambrian history of Earth's rotation and the Moon's orbit. Reviews of Geophysics, Volume 38, Issue 1, p. 37-60.

\end{thebibliography}
\end{document}